\documentclass[journal,twoside,web]{ieeecolor}
\usepackage{jsen}
\usepackage{cite}

\usepackage{amsmath,amssymb,amsfonts}
\usepackage{algorithmic}
\usepackage{graphicx}
\usepackage{textcomp}
\usepackage{wrapfig}
\usepackage{amsthm}
\usepackage[english]{babel}
\usepackage{multirow}
\usepackage{esint}
\usepackage{algorithm} 
\usepackage{algorithmic} 
\usepackage{subfig}
\usepackage{caption}
\usepackage{lipsum}
\usepackage{multicol}
\usepackage{comment}
\usepackage{epstopdf}

\def\BibTeX{{\rm B\kern-.05em{\sc i\kern-.025em b}\kern-.08em
    T\kern-.1667em\lower.7ex\hbox{E}\kern-.125emX}}
\definecolor{abstractbg}{rgb}{0.89804,0.94510,0.83137}
\begin{document}
\title{An Adaptive Video Acquisition Scheme for Object Tracking and its Performance Optimization}
\author{Srutarshi Banerjee, Henry H. Chopp, Juan G. Serra, Hao Tian Yang, Oliver Cossairt and A. K. Katsaggelos
\thanks{Manuscript submitted on February 19, 2021. \emph{(Corresponding author: Srutarshi Banerjee.)}}
\thanks{Preliminary work has been presented in \cite{PrevWork}. This work is supported in part by a the Defense Advanced Research Projects Agency (DARPA) under Grant No. HR0011-17-2-0044.}
\thanks{Srutarshi Banerjee, Henry H. Chopp, A. K. Katsaggelos are with the Department of Electrical and Computer Engineering (ECE), Northwestern University (NU), Evanston, IL, 60208, USA. (email: srutarshibanerjee2022@u.northwestern.edu; henrychopp2017@u.northwestern.edu; a-katsaggelos@northwestern.edu ).}
\thanks{Juan G. Serra is with the Center for Interdisciplinary Exploration and Research in Astrophysics, NU, Evanston, IL, 60208, USA. (email: jgserra@northwestern.edu).}
\thanks{Hao Tian Yang was with the Computer Science (CS) Department, NU, Evanston, IL, 60208, USA. (email: ThomasYang@u.northwestern.edu).}
\thanks{Oliver Cossairt is with the Department of CS and Department of ECE, NU, Evanston, IL, 60208, USA. (email: oliver.cossairt@northwestern.edu).}}

\IEEEtitleabstractindextext{%
\fcolorbox{abstractbg}{abstractbg}{%
\begin{minipage}{\textwidth}%
\begin{wrapfigure}[12]{r}{3in}%
\includegraphics[width=3in, height=1.5in]{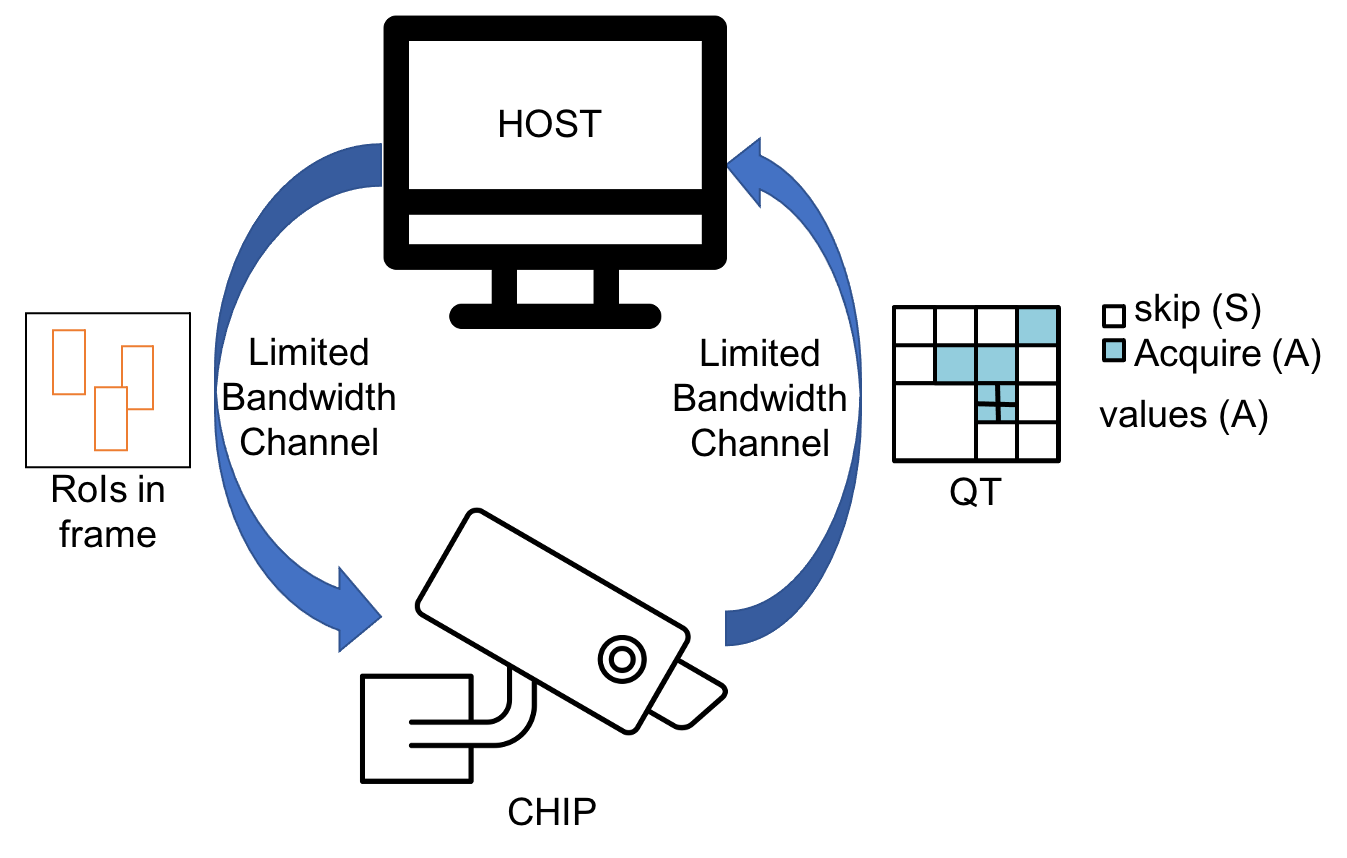}%
\end{wrapfigure}%
\begin{abstract}
We present a novel adaptive host-chip modular architecture for video acquisition to optimize an overall objective task constrained under a given bit rate. The chip is a high resolution imaging sensor such as gigapixel focal plane array (FPA) with low computational power deployed on the field remotely, while the host is a server with high computational power. The communication channel data bandwidth between the chip and host is constrained to accommodate transfer of all captured data from the chip. The host performs objective task specific computations and also intelligently guides the chip to optimize (compress) the data sent to host. This proposed system is modular and highly versatile in terms of flexibility in re-orienting the objective task. In this work, object tracking is the objective task. While our architecture supports any form of compression/distortion, in this paper we use quadtree (QT)-segmented video frames. We use Viterbi (Dynamic Programming) algorithm to minimize the area normalized weighted rate-distortion allocation of resources. The host receives only these degraded frames for analysis. An object detector is used to detect objects, and a Kalman Filter based tracker is used to track those objects. Evaluation of system performance is done in terms of Multiple Object Tracking Accuracy (MOTA) metric. In this proposed novel architecture, performance gains in MOTA is obtained by twice training the object detector with different system generated distortions as a novel 2-step process. Additionally, object detector is assisted by tracker to upscore the region proposals in the detector to further improve the performance.
\end{abstract}

\end{minipage}}}

\maketitle

\section{Introduction}
\label{sec:introduction}
\IEEEPARstart{T}{his} work focuses on the problem of optimal information extraction in wide-area surveillance using high resolution sensors with low computational power for imaging applications. The imaging instrument (i.e., the chip) is assumed to be of a very high resolution Focal Plane Array (FPA) (e.g., $> 250$ MPixels) \cite{giga1}, \cite{giga2}, \cite{giga3} providing imagery over desired field of view, but with low computational power. Imagers of such high resolution capture data at a large bit rate, but do not process them fast enough. Limited computational power in FPAs and other imaging devices is a key practical constraint in the devices currently available in the market. Moreover, the FPA contains Readout Integrated Circuit (ROIC) electronics, and the primary challenge is that the data bandwidth of the ROIC limits the maximum amount of data (in bits/s) that can be delivered by the sensor (chip). For such a sensor with low computational power capturing data at a high rate, the data needs to be analyzed remotely on a server with high computational power, termed as host, in order to perform computationally heavy tasks such as object detection, tracking, anomaly detection. For a case of a very high bandwidth and high readout rate from the chip, the chip can easily send all its captured high resolution video frames to the host for data analysis, and the analysis of the data on the host can be straight-forward with state-of-art algorithms. However, in practice, having a very high data bandwidth is impractical due to various factors: ROIC electronics, commercial aspects to using large data bandwidth, lossy transmission media and other factors. Thus, the chip can only send limited data to the host. In such a scenario, the chip must be selective in sending a subset or a compressed representation of the captured high resolution video frames. Optimally selecting the compressed video frames is a challenging task for the chip. Moreover, the host has access to only the compressed frames. Task specific computations (such as object detection, tracking) are difficult to be performed on compressed frames than high quality frames.

Commercial FPAs have different controls over spatio-temporal sampling. Pixel-binning and sub-sampling modes allow a dynamic trade-off between spatial and temporal resolutions. For instance, high frame rates (e.g., $> 1$ kfps) may be achieved at low resolution (e.g., $<$ VGA), while maximum frame rates that can be achieved for high resolution FPAs (e.g., $> 10$ MPixels) are typically low ($< 60$ Hz). The pixel binning and sub-sampling modes provide a way to optimize sampling with constraints on the bandwidth of ROIC electronics \cite{ROIC4} - \cite{ROIC9}. Analysis of theoretical and experimental properties of compressive video cameras which utilize per-pixel coded exposure sequences has been done as well \cite{ROIC4} - \cite{ROIC6}. FPA implementations using programmable coded exposure sequences have also been developed \cite{ROIC10}, \cite{ROIC11}. All of these algorithms work with an approach to maximize the information content in space and time using a compressed sensing approach which assumes the signal can be sparsely represented in a transform domain. Unfortunately this assumption is not always valid for arbitrary scenes.

We propose an architecture which performs not only the objective task (such as object detection and tracking) but also an intelligent system which can adapt its acquisition based on the scene. In order to do so, we use object detection and tracking algorithms on the host which has high computational power to perform such tasks at low computational time. Object detection and tracking has been a topic of immense interest for the image processing community over the past decade, accelerated with the advent of Deep Learning approaches \cite{AlexNet17}, \cite{Hinton18} especially with Convolutional Neural Network (CNNs), which is one of the most widely used models of deep learning  \cite{ObjDetReview}, \cite{ObjDetReview1}, \cite{ObjTrackReview}.

In this work, we introduce an algorithm for adaptive sampling of high bit rate data (such as from a high resolution FPA) that is optimized together with a reconstruction algorithm for object detection and tracking purposes. We develop the architecture assuming the imaging device (chip) to have limited computational power and the host to have high computational power. The communication channel between the chip and the host has limited bandwidth and hence, it is not possible to transfer all the captured data from the chip to host. To the best of the authors' knowledge this is the first work to introduce a bandwidth limited resource constrained optimized solution for object tracking, with our preliminary work presented in [1]. The detection and tracking of multiple objects in a compressed image domain is a unique approach in our system which requires careful optimization. Since, the framework is aimed at object tracking, the final evaluation metric for the performance of this algorithm is not the traditional reconstructed image quality measured, for example, by PSNR or SSIM, but rather a surrogate tracking performance metric, Multiple Object Tracking Accuracy, (MOTA) for tracking the objects of interest. 

The proposed host-chip architecture allows dynamic, modular, re-configurable and content-adaptive acquisition of data from an imager (chip) with low computational power, with an optimal bandwidth utilization. The optimization problem is posed as a resource allocation problem: given the constrained allowable data bandwidth between the host computer and chip, with low computational power on the chip, we estimate the best tessellation per frame based on the ROIs. A frame thus requires reduced number of bits for its representation. The host and chip mutually transmit only the most important information. The main contributions of the paper are as follows:

\begin{enumerate}
    \item The development of a host-chip modular feedback architecture designed for optimal information extraction in a bandwidth-constrained communication channel for object detection and tracking in a computationally constrained chip deployed on the field. 
    
    \item Implementation of a multiple object detector and tracker in a lossy image domain.
    
    \item Performance optimization of the proposed system using a 2-step object detector training strategy.
    
    \item Performance optimization of the proposed system by enhancing object detection with assist of an object tracker.
    
    \item Performance comparison of the proposed system with state-of-art systems.
\end{enumerate}

One of the shortcomings of this approach is the fact that the object tracking metric is used as performance tracking metric, instead of the traditional PSNR or SSIM metric. For highly constrained bandwidth, the quality of image frames might be have large distortion, while the object tracking performance may be good. In such cases, the priority is given to achieving a good object tracking metric instead of a good image quality. Our architecture focuses on having a good tracking metric for highly distorted frames. The rest of the paper is organized as follows: in Section II and III, we describe the related work and problem formulation respectively. Section IV describes the host-chip architecture. Section V describes the performance optimization of the system using a 2-step training approach for the object detector, as well as tracker assisted object detection. Section VI describes our experimental results. Section VII covers the discussion, while Section VIII concludes the paper.

\section{Related Work}

Image acquisition with adaptivity has been introduced in several ways. For example, local features e.g., standard deviation \cite{STDV12}, \cite{STDV13}, edge counting \cite{EC14} or estimation of the reconstruction error \cite{RECE15} in local  domain is used to guide the adaptive acquisition. An adaptive scheme proposed in \cite{AdapSc16} by estimating the compression based on local redundancy measured statistically utilizing previously sensed measurements. 

In the literature of rate-distortion for video / images, one of the early works for region based rate control for H.264 \cite{h_264} has been done in \cite{Rate_Region_Control}. The work used region-based rate control scheme for macro-blocks and grouped regions of similar characteristics as same region for treating them as a basic unit for rate control. Similar works on region-classification-based rate control for Coding Tree Units (CTUs) in I-frames to improve reconstruction quality of I-frames for suppressing flicker artifacts has been done in \cite{Region_class_I_frames}, region-based inter-frame rate-control scheme to improve the objective quality and reduce PSNR fluctuations among CTUs \cite{Region_based_CTU}, moving regions have been used as the RoI for identifying the depth level of CTU \cite{MR_RoI_CTU}, and other works \cite{hu2017region}, \cite{pham2016efficient}, \cite{wang2018region}. Further progress has been made in the RoI aware rate-control where higher bit rate is allocated to regions of interest human faces \cite{li2016region} and combination of human faces with CTU level \cite{meddeb2014region} and using human faces and tile-based rate control \cite{meddeb2015icip}. Work has been done in attention region based rate control for 3DVC depth map coding based on regions classified as foreground, edges of objects and dynamic regions \cite{lee2016attention}. A content-aware rate control scheme for HEVC \cite{HEVC} has been done on static and dynamic saliency detection using deep convolutional network for extracting static saliency map \cite{sun2020content}. Work has been done in preserving scale-invariant features such as SIFT/SURF \cite{SURF_preserve}. The rate control algorithm of HEVC based on RoI using improved Itti algorithm \cite{song2020optimized} has been done. Rate control has been also done by $\lambda$ adjustment in \cite{rate_control_lambda} and in other works. 

While some of these works had developed compression algorithms based on priority regions, none of them focused on joint Rate Distortion optimization with object tracking as an end metric. Moreover, these works in literature have been developed without considering computational power of the imager. None of them have a host-chip architecture with shared computational load on the chip and host, with the host performing heavy computation and providing feedback to the chip having low computational power to adaptively acquire data for the next time instant.

Object detection using Deep Neural Network has been a topic of heavy activity in the last few years. Typically, CNNs have deeper architectures which allows hierarchical feature representation and learning with fewer weight parameters, which increases their expressive capability compared to shallow models \cite{LeCunn_DLR}. R-CNN \cite{CNN1}, Spatial pyramid pooling (SPP)-net \cite{SPPNet}, Faster R-CNN \cite{CNN3}, Mask R-CNN \cite{MaskRCNN}, Region-based fully convolutional network (R-FCN) \cite{RFCN}, G-CNN \cite{GCNN}, MultiBox \cite{MultiBox}, YOLOv3 \cite{YOLOv3}, YOLOv4 \cite{YOLOv4}, Single Shot MultiBox Detector (SSD) \cite{SSD}, Deconvolutional Single Shot Detector (DSSD) \cite{DSSD} are few of the deep learning based object detectors in literature. Typically, object detectors are trained on datasets such as COCO \cite{COCO}, PASCAL VOC \cite{VOC} and ImageNet \cite{ILSVRC} which have low inherent distortions and noise. This breed of trained object detectors is not optimized for distorted frames and leads to sub-optimal detection performance. We address this challenge by retraining the object detector at different distortions to handle degraded data and boost its performance.

Object Tracking, such as the multiple object tracking (MOT) problem is typically solved using Joint Probabilistic Data Association (JPDA) filters \cite{JPDA1}, \cite{JPDA2} or Multiple Hypothesis Tracking (MHT) \cite{MHT1}. Real time applications of these approaches in highly dynamic environments is impractical due to their complexities. Online trackers build appearance model of individual objects \cite{Tracker1} - \cite{Tracker3} or a global model \cite{Tracker4} - \cite{Tracker6}. Often motion is taken into account in addition to appearance model \cite{Tracker7}. Geiger et. al. \cite{Tracker8} used the Hungarian algorithm \cite{Tracker9} in a two-stage process by forming tracklets by associating detections and then associating tracklets to bridge broken trajectories. Bewley et. al. \cite{Tracker10} used a Kalman filter based approach combined with the Hungarian algorithm in order to track multiple objects. More recently deep learning based multi-object trackers has been introduced such as \cite{deepsort} - \cite{fang2018}. While advances in object detection and tracker is still an interesting pursuit, in this work we focus on introducing a novel modular host-chip architecture which has the flexibility of upgrading the object detector and tracker with the progress of research in that domain. Typically, the object detector independently detects objects in a frame. However, we utilize the predicted locations of the object in a frame by the tracker to guide the object detections.

\section{Problem Formulation}
Our method is based on a computational imaging approach using a prediction-correction feedback paradigm. The goal of the host computer is to predict the location of the regions of interests (RoIs) for a particular frame and be able to correct that prediction. The predicted ROIs for the chip, consisting of the FPA and ROIC, help guide the chip to capture optimal information for the host to optimally perform object detection and tracking. The methodology has been developed with consideration of limited computational power on the chip which forces it to transfer data to the host to perform heavy computations.

The adaptive segmentation is data-driven based on a decomposition of the image into regions or blocks. While our architecture supports different distortion/compression introduced by these regions/blocks, in this work we focus on adaptive segmentation of video frame based on a quadtree (QT) structure. We particularly use the QT structure as this fits into the H.264 \cite{h_264}, H.265 / High Efficiency Video Coding (HEVC) \cite{HEVC} and latest H.266 / Variable Video Coding (VVC) \cite{VVC} standards which partitions image frame into QT blocks. Thus, our architecture can be applied directly into the existing electronic hardware systems which utilizes latest HEVC or VVC standards and earlier H.264 standards as well. 

The host-chip system has been developed as a prediction-correction feedback system as shown in Fig. 1. The host predicts the RoIs in a frame and updates its prediction based on the data received from the chip. This feedback mechanism is very critical for our system as it prevents error propagation. The chip generates an optimized QT structure that subdivides the current frame into superpixels before transmitting them to the host. The bigger superpixels have high distortion which may be mitigated by subdividing them if sufficient bandwidth is available. Further QT subdivision, depending on available bandwidth, captures finer details in a frame. QTs for newly acquired frame on the chip contains information about the superpixels the host should update or skip in its frame of the previous time step. The intensities for the update regions are sent from the chip to the host. Skipped superpixels assume the value of the previous frame. The QT is optimized based on: (i) the distortion between the current and previously reconstructed frame, (ii) the predicted locations of the ROIs for the current frame, and (iii) the available bandwidth. In this work, a fast and effective recursive encoding of the QT structure in \cite{Schuster1} is used. 

\begin{figure}[htbp]
\centerline{\includegraphics [width=250pt]{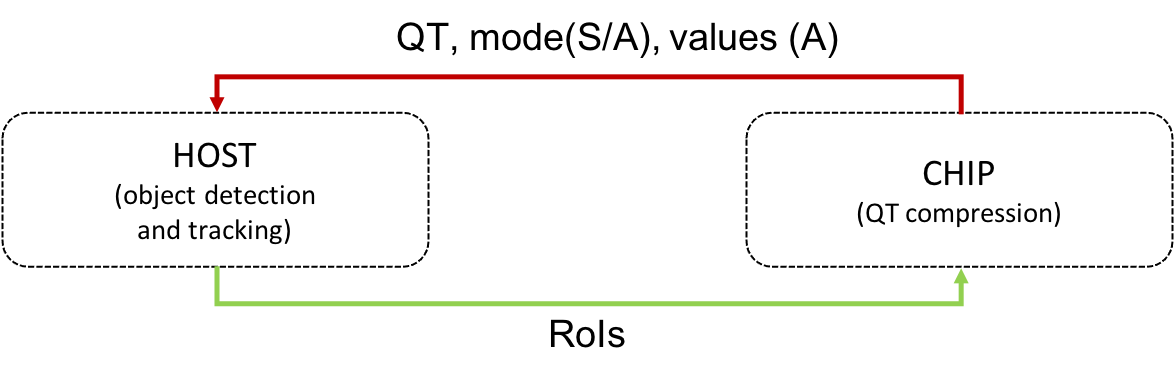}}
\caption{Host-Chip Architecture (S: Skip, A: Acquire).}
\label{fig_1}
\end{figure}

The Host-Chip architecture of the system is shown in Fig. 1. The Chip for a particular frame, sends QT, mode of leaves (skip or acquire) and pixel values corresponding to the acquire mode to the Host. The Host based on these information computes the ROIs for the next frame and sends in back to the chip. This iterative loop is repeated once for each frame the chip captures. Clearly, the host has access to only distorted frames which are compressed by the QT. The object detector on the host needs to classify and return bounding boxes based on these distorted frames, which is more challenging compared to the undistorted, higher quality frames. The performance of the object detector deteriorates due to the QT compression, and hence it is a necessity to boost its performance under low bandwidth conditions. This is of utmost importance for the host-chip architecture which must be robust to both bandwidth fluctuations and different operating conditions. Additionally the object detector uses spatial information per frame to generate bounding boxes. In order to maintain a temporal continuity among the bounding boxes, the RoIs predicted by the object tracker is taken into account. Section IV of this paper details the 2-step training process as well as the tracker assisted object detection. The performance metric for this host-chip architecture is Multiple Object Tracking Accuracy (MOTA) instead of the PSNR metric typically used in literature, as the end system performance is of critical importance, rather than the fidelity to the undistorted frames. This helps in maintaining the objective performance metric as the criteria for comparison.

\section{Host-Chip System Architecture}
The system architecture consisting of a host-chip framework is developed from the methodology of guiding a sensor (chip) through real-time tuning of its optimization parameters to collect data with highest content of useful information for object tracking. The architecture is based on the consideration of limited bandwidth channel capacity, $B$, between the host computer and chip with limited (low) computational power on the chip. The host-chip modular architecture has been developed keeping in mind the predictive-correction feedback system. The chip has low computational power while the host has high computational power. The disparity between the computational power of the chip and host drives the design of the host and chip models.

\begin{figure}[htbp]
\centerline{\includegraphics [width=250pt]{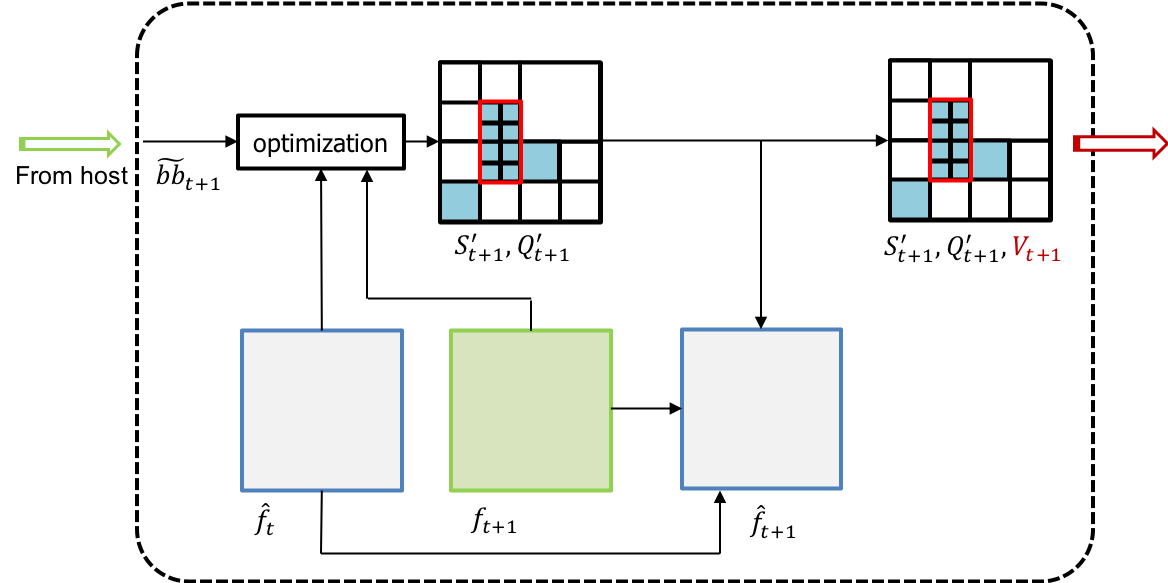}}
\caption{Computation on Chip}
\label{fig_2}
\end{figure}

\subsubsection{Chip Computation}
Fig. \ref{fig_2} shows the computation on the chip. The compression of video frame is based on a QT structure. The host computes the predicted bounding boxes $\widetilde{bb}_{t+1}$, with $\widetilde{bb}_{t+1} \in \mathcal{R}^{4 \times P}$ ($P$ is the number of bounding boxes detected), and sends it to the chip for time $t+1$. The chip has a copy of $\hat{f}_t$, which is the distorted frame for time $t$. The full resolution undistorted frame at $t+1$, $f_{t+1}$ is acquired at time $t+1$ by the FPA on the chip. These are inputs to the Viterbi Optimization Algorithm, which provides as output the optimal QT structure $S^{\prime}_{t+1}$ and optimal skip-acquire modes $Q^{\prime}_{t+1}$ subject to the communication channel bandwidth constraint $B$ for time $t+1$. The skip (S) and acquire (A) modes in $S^{\prime}_{t+1}$ identify the QT leaves (blocks) where new data need to be acquired at time $t+1$ and the remaining leaves (QT blocks) where data will be copied from frame $\hat{f}_t$. The S and A modes are included in the framework, as this allows only a reduced set of data to be sent from the chip to the host, thereby aiding in data compression significantly. Now $\{\hat{f}_t, f_{t+1}\} \in \mathcal{R}^{N_1 \times N_2}$, where ${N_1}, {N_2} = 512$ (for instance) is the resolution of the frame, $S^{\prime}_{t+1} \in \mathcal{R}^{4 \times N}$ and $Q^{\prime}_{t+1} \in \mathcal{R}^{2 \times N}$, with $N$ as the maximum depth of the QT ($N = 9$, for $N_1 = N_2 = 512$). The bounding box information $\widetilde{bb}_{t+1}$, is used to prioritize the distortion in the RoIs relative to other regions. The higher distortion in RoI regions forces the optimization algorithm to allocate more bits while performing the rate-distortion optimization. On the chip, $S^{\prime}_{t+1}$, $Q^{\prime}_{t+1}$ provides us with the QT structure alongwith the skip/acquire modes. Corresponding to the acquire modes in $Q^{\prime}_{t+1}$ and acquired frame at $t+1$, $f_{t+1}$, we can generate the pixel values for the leaves (QT blocks), $V_{t+1}$ for the acquire modes. Here, $V_{t+1} \in \mathcal{R}^{N_a}$, with $N_a$ as the number of acquire modes in $Q^{\prime}_{t+1}$. The chip sends $S^{\prime}_{t+1}$, $Q^{\prime}_{t+1}$ and $V_{t+1}$ to the host in order to reconstruct the frame $\hat{f}_{t+1}$. The differential information is sent from the chip to the host, instead of the whole frame. This helps in reducing the bandwidth required for transferring the relevant information to the host for performing the tasks of object detection and tracking.

\emph{Viterbi Optimization}

The Viterbi optimization provides a trade-off between the frame distortion $D$ and frame bit rate $R$. This is done by minimizing by frame distortion $D$ over the leaves of the QT $\textbf{x}$ subject to a given maximum frame bit rate $R_{max}$. Here, $\{D, R\} \in \mathcal{R}^{4 \times N}$, $\textbf{x} \in \mathcal{R}^{4 \times N}$ and $R_{max} \in \mathcal{R}$, where $N$ is the maximum depth of the QT. Previous works \cite{Schuster1}, \cite{Schuster2}, \cite{AKK1} on Viterbi optimization have been used for compression on actual frames. In this work, we use the reconstructed frame $\hat{f}_{t}$ and the actual frame $f_{t+1}$ acquired by the chip to compute the distortion.

The optimization is formulated as follows
\begin{align}
\label{eqn:eqlabel1}
   \arg\min_{\textbf{x}} & {~D(\textbf{x})}, \\
   \text{s. t. } & {R(\textbf{x}) \leq R_{max}} \nonumber
\end{align}

The distortion for each node of the QT is based on the acquisition mode $Q^{\prime}_{t+1}$ of that node. If a particular node $\hat{x}_t$ of a reconstructed frame at time $t$ is skip, the distortion with respect to the new node at time $t+1$, $x_{t+1}$, is given by

\begin{align}
\label{eqn:eqlabel2}
    D_s = | x_{t+1} - \hat{x}_t |,
\end{align}

On the contrary, if the node is an acquire, the distortion is proportional to the standard deviation $\sigma$. This is shown in Eq. \ref{eqn:eqlabel3}, where $N$ is the maximum depth of the QT and $n$ is the level of the QT where distortion is computed. The root is defined to be on level $0$, and the most subdivided level as $N$:

\begin{align}
\label{eqn:eqlabel3}
    D_a = \sigma \times 4^{N-n},
\end{align}

It must be kept in mind that the distortion $D$ is computed per block the QT and thus $\{D_s, D_a\} \in \mathcal{R}$. The total distortion is therefore defined as

\begin{align}
\label{eqn:eqlabel4}
    D = D_s + D_a,
\end{align}

\begin{figure}[htbp]
\centerline{\includegraphics [width=240pt]{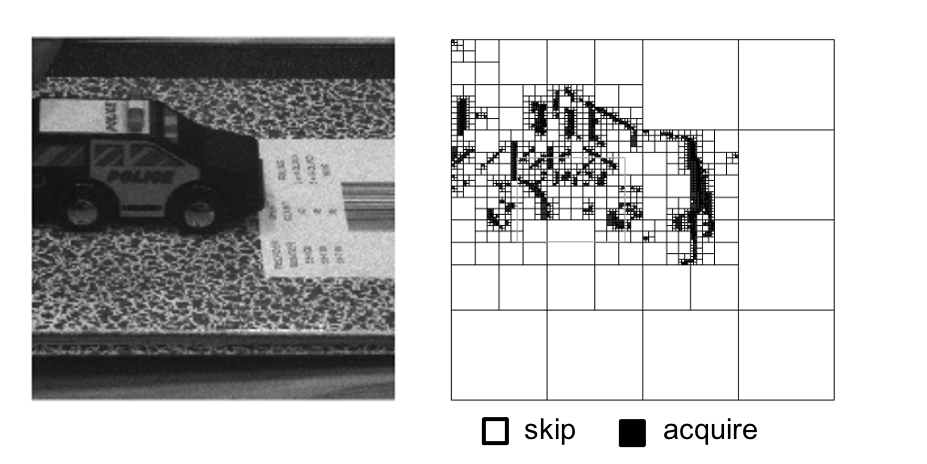}}
\caption{Sample frame with (left) with its QT decomposition (right) for $\lambda = 2.5$.}
\label{fig_sample_fr}
\end{figure}

The constrained discrete optimization of Eq. \ref{eqn:eqlabel1} is solved using Lagrangian relaxation, leading to solutions in the convex hull of the rate-distortion curve \cite{Schuster2}. The Lagrangian cost function is of the form

\begin{align}
\label{eqn:eqlabel5}
    J_{\lambda} (\textbf{x}) = D(\textbf{x})+
    \lambda R(\textbf{x}),
\end{align}
where $\lambda \geq 0$, ($\lambda \in \mathcal{R}$) is a Lagrangian multiplier. Here, $J_{\lambda}(\textbf{x}) \in \mathcal{R}^{4 \times N}$, over all the leaves of the QT. It has been shown that if there is a $\lambda^*$ such that

\begin{align}
\label{eqn:eqlabel6}
   \textbf{x}^* = \arg\min_{\textbf{x}} & { J_{\lambda^*}(\textbf{x})}
\end{align}
which leads to $R(\textbf{x}^*) = R_{max}$, then $\textbf{x}^*$ is the optimal solution to Eq. \ref{eqn:eqlabel1}. This is solved using the Viterbi algorithm, shown in detail in \cite{Schuster1}. A sample frame with its QT decomposition containing the skip and acquire modes are shown in Fig. \ref{fig_sample_fr} for $\lambda = 2.5$ which corresponds to regime of low distortion.

Now, in the distortion term, we want to prioritize the regions based on the bounding boxes, which are the ROIs of region $i$. This is introduced by the weight factors $w_i$ in each region $i$. However, in case where region $i$ occupies a large area within the frame, the amount of distortion may heavily outweigh other smaller regions. We want to have a weighted distortion independently of the area of ROI $i$. This is done by dividing the weighted distortion by the area of the ROI of region $i$, thus modifying Eq. \ref{eqn:eqlabel5} as 

\begin{align}
\label{eqn:eqlabel7}
    J_{\lambda} (\textbf{x}) =  \sum_{i\in\Omega}\frac{{w_i}{D_i(\textbf{x}_i)}}{A_i} + 
    \lambda R(\textbf{x}),
\end{align}
where, 
$\Omega$ is the set of differently weighted regions, 
${D_i}$ the distortion of region $i$ ($D_i \in \mathcal{R}$), 
$w_i$ the weight of region $i$ ($w_i \in \mathcal{R}$), 
$A_i$ the area of region $i$ ($A_i \in \mathcal{R}$), and
$\textbf{x}_i$ the leaves in the QT of region $i$.

The system can also be operated in a fixed bit rate within a certain tolerance. The $\lambda$ value in the Lagrangian multiplier is adjusted at each frame for achieving the desired bit rate. The optimal $\lambda^{*}$ is computed by a convex search in the Bezier curve \cite{Schuster2}. The Bezier curve accelerates convergence in fewer iterations.

\subsubsection{Host Computation}

\begin{figure}[htbp]
\centerline{\includegraphics [width=250pt]{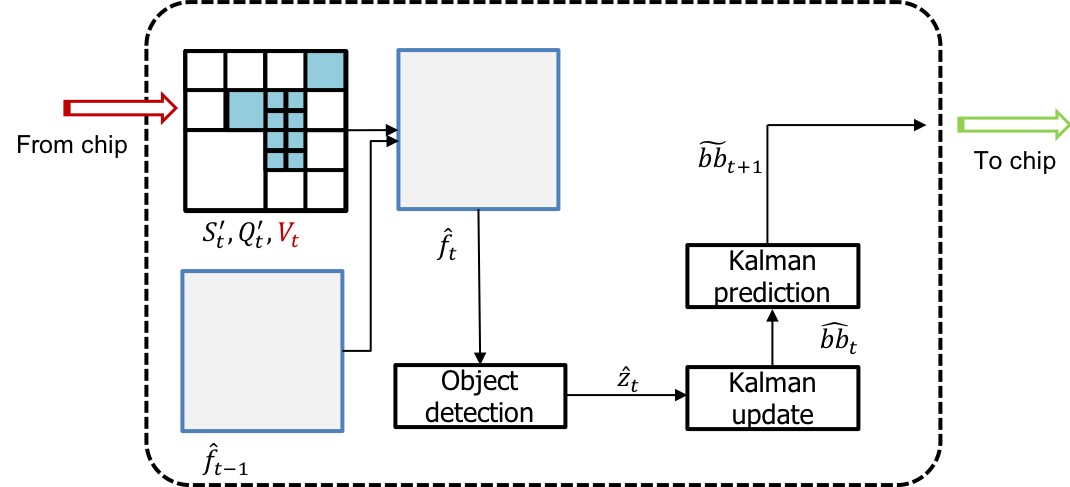}}
\caption{Computation on Host.}
\label{fig_3}
\end{figure}

Fig. \ref{fig_3} shows the computation on the host. For a undistorted frame $f_t$ acquired at time $t$ on the chip, we have QT acquisition, skip or acquire modes for the leaves, and values for the leaves of acquire modes, denoted by ${S}^{\prime}_t$, $Q^{\prime}_t$, and $V_t$, respectively. These are then sent from the chip to the host in order reconstruct frame $\hat{f}_t$. 
The previously reconstructed frame $\hat{f}_{t-1}$ for time $t-1$ saved on the host is used to copy the values in the skip leaves of $\hat{f}_t$. Here $\{f_t, \hat{f}_t, \hat{f}_{t-1}\} \in \mathcal{R}^{N_1 \times N_2}$, where ${N_1}, {N_2} = 512$ (for instance) is the resolution of the frame, $S^{\prime}_t \in \mathcal{R}^{4 \times N}$ and $Q^{\prime}_t \in \mathcal{R}^{2 \times N}$, with $N$ as the maximum depth of the QT ($N = 9$, for $N_1 = N_2 = 512$). An object detector on the host then determines the ROIs of the reconstructed image $\hat{bb}_t$. The ROIs are then fed into a Kalman Filter-based object tracker as an observation, which updates the state of the filter. The Kalman Filter then predicts the locations of the next ROIs for the next frame at time $t+1$, based on a linear motion model, denoted as $\widetilde{bb}_{t+1}$. Here, $\{\hat{bb}_t, \widetilde{bb}_{t+1}\} \in \mathcal{R}^{4 \times P}$ ($P$ is the number of bounding boxes detected). These predicted ROIs for frame at $t+1$ are then sent back to the chip. A copy of the distorted reconstructed frame $\hat{f}_{t}$ is kept in the host for creating the reconstructed frame $\hat{f}_{t+1}$ at time $t+1$.

\emph{Object Detection}

The regions of interest is detected by using an object detector based on the reconstructed frame on the host as shown in Fig. \ref{fig_3}. While in principle the framework supports any object detector, in this work, we use Faster R-CNN \cite{CNN3} for detecting objects of interest owing to its higher accuracy than other deep learning based object detectors. Infact, Faster R-CNN is widely used in several systems. Faster R-CNN comprises of two modules: the first module consists of the convolutional layers of VGG16 \cite{VGG16} which extracts features. A region proposal network (RPN) finds and labels regions of probable objects as foreground or background. The second module classifies the objects in those region proposals and also regresses a bounding box for each object. This object detector on the host has access to only distorted reconstructed frames. For enhancing its performance on degraded data as well, the object detector has been trained on distorted and undistorted data. Additionally, in order to ensure continuity among the frames in terms of detected objects, the bounding boxes predicted by the tracker is used to assist the Faster R-CNN. Multiple classes of objects were used to train the Faster R-CNN network. In this work, we train the object detector using a novel 2-step methodology and assisted by tracker information as described in Section \ref{Perf_Opt}.

\emph{Object Tracker}

The object detector generates bounding box with class labels, which are fed as input to an object tracker. While in principle the framework supports any tracker, in this work, a Kalman Filter-based multiple object tracker, Simple Online and Realtime Tracking (SORT) \cite{Tracker10} is adapted in this object tracker implementation. The object tracker uses a linear motion model to predict the bounding box locations in the next frame $f_{t+1}$. It then associates the identities using linear assignment between the new detections from Faster R-CNN and the most recently predicted bounding boxes. The state of the Kalman Filter, $\textbf{X}_s$, for each detection is modeled using a linear motion model as

\begin{equation}
   \textbf{X}_s = [u,v,s,r,\dot{u},\dot{v},\dot{s}]^T,
\end{equation}
where $u$ and $v$ represent the coordinates of the target's center, and $s$ and $r$ represent the scale (area) and the aspect ratio (width/height) of the target’s bounding box, respectively. Three of these time derivatives are part of the state parameters as well, namely $\dot{u}$, $\dot{v}$, and $\dot{s}$.

When a detection is associated with a target, the target state is updated using the detected bounding box. The velocity components of the state are solved optimally via the Kalman filter framework \cite{KF}. The predicted bounding boxes are extracted from predicted state of the Kalman filter. These are the ROIs for acquisition of the next frame $f_{t+1}$ which are also input to the Viterbi algorithm. However, when there is no detection from the object detector, the predicted bounding boxes are translated following the constant motion model for $N_{tracked}$ consecutive frames. The predicted bounding boxes are fed into the Faster R-CNN for upscoring those predictions. Additionally, the predicted regions are of higher quality due to lower distortion in those regions as described in Eqn. $7$. This allows the Faster R-CNN to detect objects in one out of $N$ frames and still be tracked using the Kalman Filte, thereby improving the tracking accuracy.

\subsubsection{Performance Accuracy Metric}
The multi-target performance is measured using the Multiple Object Tracking Accuracy (MOTA) evaluation metric defined as \cite{mota} refered here as $MOTA_{full}$,

\begin{align}
\label{eqn:eqlabel}
MOTA_{full} = 1 - \sum_{t} {\dfrac{m_t + fp_t + mme_t}{g_t}},
\end{align}
where $m_t$ represents the number of missed detections at time $t$,
${fp}_t$ the number of false positives at time $t$,
$mme_t$ the number of mismatch (track switching) errors at time $t$ and
$g_t$ the number of ground truth objects at time $t$.

In this work we also consider a modified MOTA metric which does not penalize the false positives. It is of utmost importance for many object tracking applications (including ours) that all objects that should be tracked are indeed tracked, especially when there is an increased difficulty in detecting the objects in degraded frames. The modified MOTA ($MOTA_{mod}$) is given by 

\begin{align}
\label{eqn:eqlabel}
MOTA_{mod} = 1 - \sum_{t} {\dfrac{m_t + mme_t}{g_t}},
\end{align}

A higher score of $MOTA_{mod}$ and $MOTA_{full}$ corresponds to higher tracking of the objects in the video sequence and hence better performance. The experiments are conducted for different values of $\lambda$ in reference to Eq. \ref{eqn:eqlabel5}, which provides operating point in the rate-distortion curve. This provides different average bit rates over a video sequence, which are a fraction of the maximum rate. For different values of $\lambda$, the distortion and the bit rate fluctuates for each frame. However, in practice the communication channel between the chip and the host is bandwidth-limited. Thus the bit rate of the data sent through the channel can only vary within a certain tolerance (e.g., $< 1 \% $). In order to keep the bit rate constant, for each frame we vary $\lambda$. This mode of operation keeps the rate fixed, within certain tolerance, but the distortion varies frame to frame.

\section{Performance Optimization}\label{Perf_Opt}
The system is designed to achieve good object tracking performance for different bit rates $R$. The object detector identifies the ROIs, which are then input to the object tracker. Hence, it is the most important component of the host in its role of detecting and tracking objects in each frame. However, the host has access to only the reconstructed frame $\hat{f}_t$ at time $t$, which is a distorted version of the uncompressed high quality frame. In order to perform well the Faster R-CNN must also be trained with similarly distorted frames. This would improve the detection accuracy of the Faster R-CNN for system-generated distortions at different bit rates.

\subsection{Training the Object Detector}
Traditionally, object detectors are trained on data from publicly available datasets such as COCO \cite{COCO}, PASCAL VOC \cite{VOC} and ImageNet \cite{ILSVRC}, among others. Most of the datasets have been curated using a good quality camera, and the inherent distortions and noise in those image/video frames is low. Thus, these object detectors are finely tuned to the image quality of the particular dataset. The detection performance worsens once it is tested with other forms of distortion. In order to address this issue and improve the performance of the detector on distorted frames, we resort to training the object detector in a novel two stage approach. This two step approach achieves much higher performance with system-generated distortions than training with undistorted images. We used the ILSVRC VID dataset \cite{ILSVRC} to train the Faster R-CNN. Since the work is catered to surveillance applications in ground, air and water scenes, we trained our object detector on Airplanes, Watercrafts and Cars. However, it must be kept in mind that the architecture can work with an object detector trained on any number of classes. The training data in this dataset has been split into 70:30 randomly as training and validation data for training the Faster R-CNN.

\subsubsection{Step I}
In this step, the object detector in the host in Fig. \ref{fig_3} is replaced by Ground Truth bounding boxes. This creates exact bounding boxes (ROIs) precisely encompassing the entire object while still generating data consistent with the degradation we would see in the system.

\begin{comment}
\begin{figure}[htbp]
\centerline{\includegraphics [width=250]{Fig3a.png}}
\caption{Step-I Training (Object Detector on Host (Fig. \ref{fig_3}) replaced by Ground Truth.}
\label{fig_4}
\end{figure}
\end{comment}

The ROIs are then transmitted to the chip. The chip finds the optimal QT according to the ROIs, $\lambda \in \{50, 100, 250, 400, 650\}$ (the value in the Viterbi optimization algorithm), along with the full undistorted frame $f_t$ on the chip and the previous reconstructed frame $\hat{f}_{t-1}$. The distortion levels in the system are set by the weights $w_i$ in the ROIs and background. The weights are uniquely selected such that the resulting distortion in the background is significantly higher than that in the ROIs. For each value of $\lambda$, the entire training data is passed through the architecture which from $\hat{f}_t$ creates the training and validation dataset for Faster R-CNN. The data in the original dataset corresponding to $\lambda = 0$ is also included in the dataset. The Faster R-CNN trained on this distorted data has seen high quality data as well as data with different degrees of distortion corresponding with $\lambda$. The higher the $\lambda$ is, the higher the distortion. Ground truth annotations are used for training and validation of the Faster R-CNN.

\subsubsection{Step II}
\begin{figure}[htbp]
\centerline{\includegraphics [width=150pt, height=120pt]{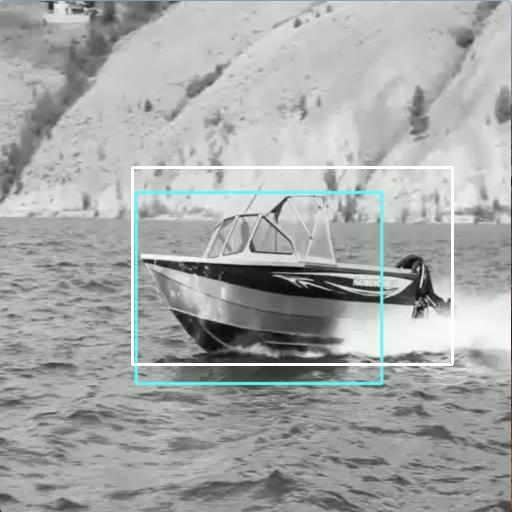}}
\caption{Bounding box and object partly overlapping.}
\label{fig_5}
\end{figure}

The Faster R-CNN trained in Step I has been trained on perfect bounding boxes which encompass the object completely. However, in actual scenarios, the object detector may detect bounding boxes which may not perfectly align with the object. For example, part of the bounding box may not entirely overlap with the object. An example is shown in Fig. \ref{fig_5}. The bounding box predicted by the object detector is shown in blue. This does not align perfectly with the ground truth bounding box which is in white. Clearly, the tail and top of the boat are not covered by the blue bounding box, whereas portions of the background in the bottom of the boat is included in the blue bounding box. 

Regardless, the Kalman Filter predicts ROIs for the next frame based on these imperfect detections. The chip then acquires the next frame based on these imperfections and sends them to the host. Portions of the object inside the ROI will be less distorted and portions outside the ROI will be highly distorted as per the weight ratio. In order to improve the object detector performance, the Faster R-CNN needs to be trained on this type of unique distortion - where part of the object is segmented finely with less distortion and the rest coarsely with high distortion. This is the objective of Step II training.

The Faster R-CNN trained from Step I is used as the object detector in the host as in Fig. \ref{fig_3}. The bounding boxes detected by the Faster R-CNN is passed to the Kalman Filter to update the state and predict the ROIs in the next frame. The chip reconstructs the frame based on these ROIs predicted by the Kalman Filter. Analogously to training in Step-I, for each value of $\lambda \in \{50, 100, 250, 400, 650\}$ along with original dataset ($\lambda = 0$), the entire training data is again passed through the architecture which creates $\hat{f}_t$, the training and validation data. The ground truth annotations are used for training and validation in this step as well. 

The Faster R-CNN trained in Step I, during the testing phase generates the bounding boxes closely aligned to the actual physical object. However, it never generates perfect bounding boxes exactly aligned to the actual physical object. The bounding box detections partially align with the actual objects in most of the cases. These bounding boxes are then passed onto Kalman Filter, which predicts the RoIs imperfectly compared to the actual object and sends them back to the chip. The reconstructed frame on the chip thus has different degrees of distortion corresponding to the entire actual physical object. The Step II training is hence critical as it trains the Faster R-CNN taking into account the different distortion levels for the object.

\begin{figure}[htbp]
\centerline{\includegraphics 
[width=150pt, height=120pt]{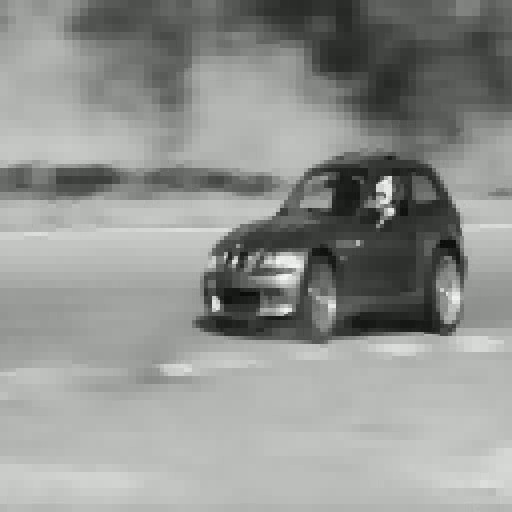}}
\caption{Example degraded frame of a car for $\lambda=100$.}
\label{fig_6}
\end{figure}

The system performance is sensitive to the training data for the object detector. The generation of distorted data for training and validating the Faster R-CNN depends on the weights assigned to the ROIs and elsewhere. This is important as it dictates the extent of relative distortion. Based on randomly selected videos from the training data, for $\lambda \in \{100, 350, 650\}$ corresponding to low, medium and high distortions respectively, we chose the weights as  $w_i = 10^7$ for the ROIs and $w_i = 10^6$ for the rest of the regions (background) with reference to Eq. \ref{eqn:eqlabel7}, which visually made distortion between the ROIs and the background distinct, with the background is not too heavily distorted compared to the ROIs. An example of such a frame is shown in Fig. \ref{fig_6}. The car within the ROI, has a finer segmentation (and therefore lower distortion) than the background.

\subsubsection{Model Variants}

We compare the tracking performance of the system with object detector models trained on different datasets. Videos including airplanes, cars and watercraft from the ILSVRC VID dataset of different distortions were used for training six different Faster R-CNN models:

\begin{enumerate}
    \item \textit{Pristine NN model}: Faster R-CNN trained exclusively with pristine (non-distorted) data
    \item \textit{Uniform NN model}: Faster R-CNN trained with pristine data and uniformly binned $2 \times 2$, $4 \times 4$ and $8 \times 8$ data
    \item \textit{Mixed NN model}: Faster R-CNN trained with pristine data and distorted data for a mixed assortment of $\lambda\in\{50, 100, 250, 400, 650\}$ generated in Step I training
    \item \textit{Mixed+ NN model}: Faster R-CNN trained with pristine data and distorted data for $\lambda\in\{50, 100, 250, 400, 650\}$ generated in Step II training
    \item \textit{MixedU NN model}: Faster R-CNN trained with pristine data, uniformly binned $2 \times 2$, $4 \times 4$ and $8 \times 8$ and distorted data for $\lambda\in\{50, 100, 250, 400, 650\}$ generated in Step I training
    \item \textit{MixedU+ NN model}: Faster R-CNN trained with pristine data, uniformly binned $2 \times 2$, $4 \times 4$ and $8 \times 8$ and distorted data for $\lambda\in\{50, 100, 250, 400, 650\}$ generated in Step II training
\end{enumerate}

In the Mixed+ model, the Mixed model is used on the Host to generate distorted data as mentioned in Step-II training. Similarly, in order to generate MixedU+ model, the MixedU model is used as the object detector to generate distorted data as mentioned in Step-II training. The NN models were trained using ADAM \cite{ADAM} as the optimizer with a learning rate of 1e-5. Dropout of 0.5 is used while training the models. During testing, no dropout is used.

\subsection{Tracker assisted Detection Framework}
The task of object tracking was initially framed to be different from object detection. However, more recent algorithms have aimed to fuse both of them together. In this work, we use the region-based object detector (Faster R-CNN) with the Kalman Filter based tracker to form a novel joint Detector-Tracker (JDT) system. This is shown in 
Fig. \ref{Thomas_overview} below. The region based object detectors (e.g. Faster R-CNN) generates lot of candidate bounding box, more than the number of objects in the scene before eventually removing most of them. To prioritise the candidate bounding boxes overlapping with the object, a set of detection confidence scores are calculated for each candidate bounding boxes. If the detection confidence score of candidate bounding boxes is lower than pre-defined threshold, those candidate bounding boxes are classified as "background" class and removed. However, this approach does not take into account any temporal continuity between the frames. 

\begin{figure*}[t]
   \centering
   \includegraphics[width=\textwidth,height=200pt]{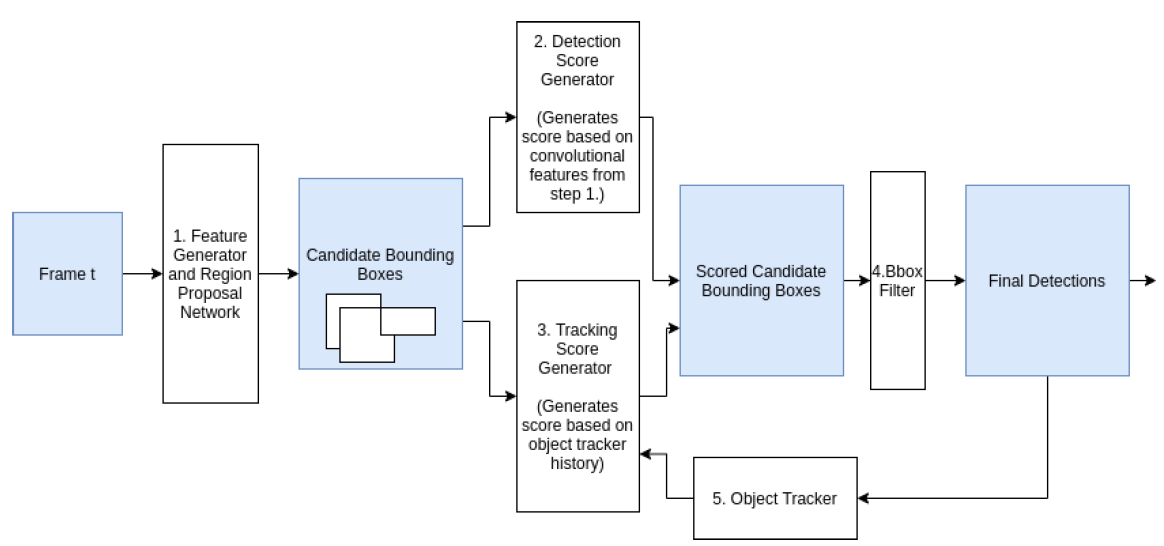}
   \caption{Tracker assisted Object Detection. The candidate bounding boxes are given detection confidence and tracking confidence scores. The tracking confidence score is updated based on predictions from object tracker. Both these scores are combined to modify the bounding box score before applying bounding box filter.}
   \label{Thomas_overview}
\end{figure*}

In order to utilize the temporal consistency among the image frames, we introduce the concept of "tracking confidence score" to describe the likelihood of a given bounding box containing a tracked object. Similar to detection confidence scores for each class of object, we introduce multiple tracking confidence scores, one for each object class. The tracking confidence scores are computed based on the highest Intersection over Union (IoU) values between all candidate bounding boxes with the bounding box predicted by the tracker. Additional constraints are forced while computing the IoU in order to remove the false positives: (1) candidate bounding boxes with IoU $< 0.3$ are rejected, and, (2) candidate bounding boxes with difference in size greater than $50 \%$ are not considered.

The joint confidence score $C_{j}$ is computed from the detection score $C_{d}$ and tracking score $C_{t}$ using Eqn. \eqref{eqn:joint_score} with $w_{t}$ and $w_{d}$ as the tunable parameter which weights the tracking confidence score and detector confidence score respectively.

\begin{align}
\label{eqn:joint_score}
C_{j} = \sqrt{w_{d}C_d^2 + w_{t}C_t^2},
\end{align}

Combining both the tracking and detection scores for the candidate bounding boxes is tricky. We fuse the two scores into a joint confidence score satisfying: (1) bounding boxes containing objects entering the scene should not have its score be penalized by lack of tracking information, (2) bounding boxes that have low detection score but high tracking score should have its joint score be boosted by virtue of its high tracking score, and, (3) bounding boxes which have mediocre detection score and tracking score should have a lower joint score than a bounding box with at least one excellent confidence score. With drop in quality of the frames, the candidate bounding boxes with low detection scores must be compensated with high tracking scores. (2) Object entering the scene without any tracking history is rewarded with higher detection or tracking score without penalizing cases where one score is much lower than other.  

\section{Experimental Results}

The experimental performance results of the system is shown in this paper by simulating the proposed model on three sequences of the ILSVRC VID dataset: (i) a video of airplanes, ILSVRC2015\_val\_00007010.mp4; (ii) a video of a watercraft, ILSVRC2015\_val\_00020006.mp4; and (iii) a video of cars, ILSVRC2015\_val\_00144000.mp4. These videos are selected to have optically small, medium, and large sized objects as well as sequences with one, two and multiple objects. The frames were resized to $512\times 512$ to accommodate the QT structure. The maximum depth of the tree is thus $N = 9$. 
 
\subsection{Variation of Distortion with Rate}
The amount of the distortions at different bit rates are important parameters in identifying the distortions which is generated by this system. We compute the variations of the distortions as PSNR and SSIM metrics for the sequences at different bit rates which is shown in Table \ref{tab:varying bit rate}. The uncompressed bit rate is 62.91 Mbits/s. The PSNR and SSIM has been computed at different $\%$ bit rate  with respect to this uncompressed bit rate. For small and medium sized objects, the PSNR and SSIM are quite low for low bit rates while for relatively larger sized objects (e.g. boat sequence), the distortions are significantly higher as shown by relatively low PSNR and SSIM values. The performance of the system has been optimized to such high distortion levels where the object is almost not recognizable. Sample frame $\hat{f}_{60}$ for different sequences at bit rate of $1.5\%$ of the maximum bit rate is shown in Fig. \ref{fig_distortions} below. Clearly the tail of the boat in the Fig. \ref{fig_distortions} (b)  is not recognizable while the cars and planes can be recognized at such bit rates as well. It should be kept in mind that the PSNR and SSIM values are only for visual quality. The end performance of the system is dictated by the MOTA metric.

\begin{table}[htbp]
    \centering
    \resizebox{\linewidth}{!}{\begin{tabular}{|c|c|c|c|}
        \hline
        Sequence & Bit Rate ($\%$) & PSNR (dB) & SSIM\\
        \hline
        \multirow{9}{*}{Airplane Sequence} & 0.75 & 26.4150 & 0.8962\\
         & 1.00 & 28.4576 & 0.9166 \\
         & 1.50 & 32.1420 & 0.9477 \\
         & 2.00 & 35.0587 & 0.9639 \\
         & 3.00 & 37.7115 & 0.9745 \\
         & 5.00 & 39.5348 & 0.9809 \\
         & 7.00 & 40.1643 & 0.9836 \\
         & 10.00 & 40.5849 & 0.9852 \\
         & 25.00 & 41.1766 & 0.9867 \\
        \hline
        \multirow{9}{*}{Boat Sequence} & 0.75 & 16.7498 & 0.4214\\
         & 1.00 & 17.9076 & 0.4411\\
         & 1.50 & 20.2070 & 0.4685\\
         & 2.00 & 21.8621 & 0.5301\\
         & 3.00 & 23.8621 & 0.5963\\
         & 5.00 & 25.0364 & 0.6713\\
         & 7.00 & 26.3222 & 0.7194\\
         & 10.00 & 27.8823 & 0.7724\\
         & 25.00 & 32.5498 & 0.8964\\
        \hline
        \multirow{9}{*}{Car Sequence} & 0.75 & 22.1733 & 0.7390 \\
         & 1.00 & 24.2945 & 0.7683 \\
         & 1.50 & 27.8283 & 0.8335 \\
         & 2.00 & 29.7073 & 0.8609 \\
         & 3.00 & 32.3193 & 0.9018 \\
         & 5.00 & 35.0724 & 0.9348 \\
         & 7.00 & 36.8468 & 0.9508 \\
         & 10.00 & 38.5461 & 0.9638 \\
         & 25.00 & 40.6701 & 0.9795 \\
        \hline
    \end{tabular}}
    \caption{SSIM and PSNR for sequences at different bit rates}
    \label{tab:varying bit rate}
\end{table}

\begin{figure*}[t]
\begin{multicols}{3}
    \noindent
    \includegraphics[width=0.8\linewidth]{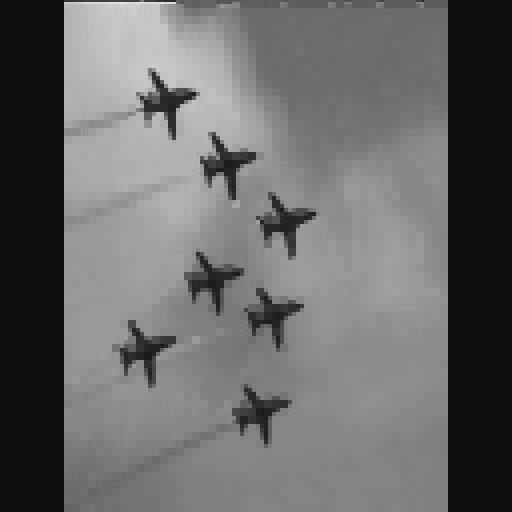}\par\caption*{(a) Airplane Sequence}
    \includegraphics[width=0.8\linewidth]{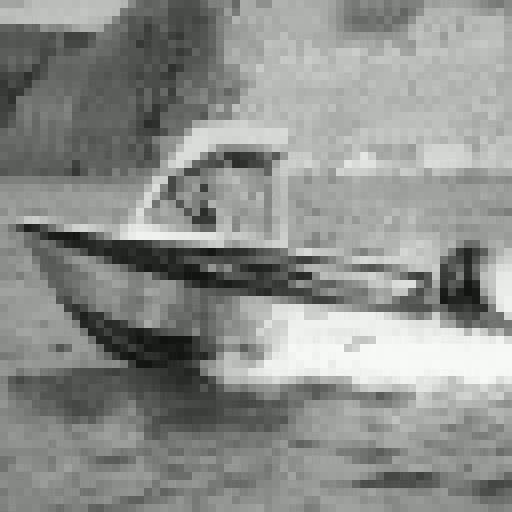}\par\caption*{(b) Watercraft Sequence}
    \includegraphics[width=0.8\linewidth]{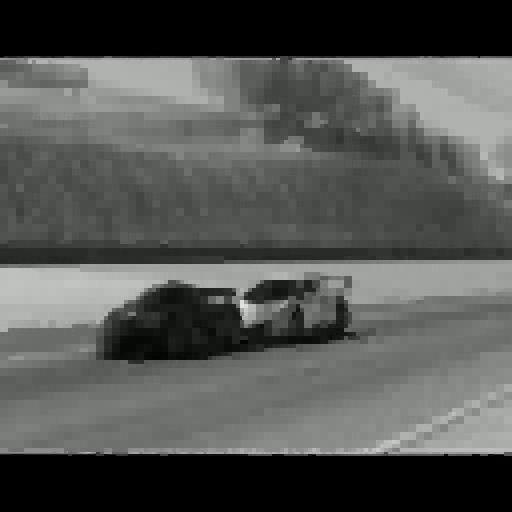}\par\caption*{(c) Car Sequence}
\end{multicols}
\caption{Degraded frame $\hat{f}_{60}$ generated at $1.5\%$ of maximum bit rate. The background has more distortion than the objects itself. We ask the reader to zoom in on each of the frames to see the degradations.}
\label{fig_distortions}
\end{figure*}

\subsection{Operation at constant $\lambda$}
\begin{figure*}[t]
\begin{multicols}{6}
    \noindent
    \includegraphics[width=\linewidth]{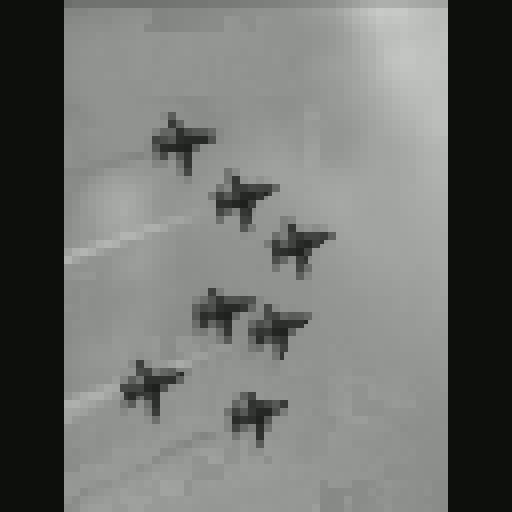}\par\caption*{Pristine NN}
    \includegraphics[width=\linewidth]{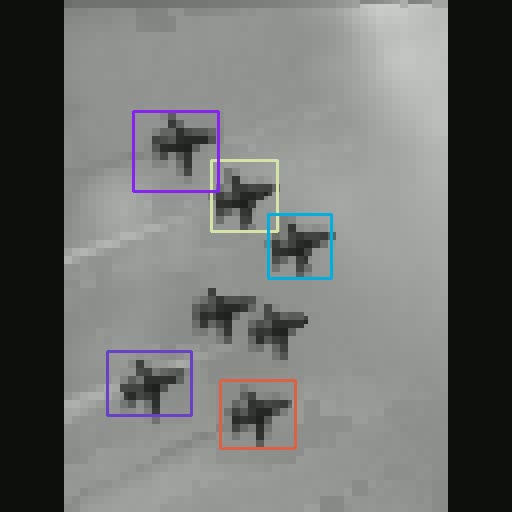}\par\caption*{Uniform NN}
    \includegraphics[width=\linewidth]{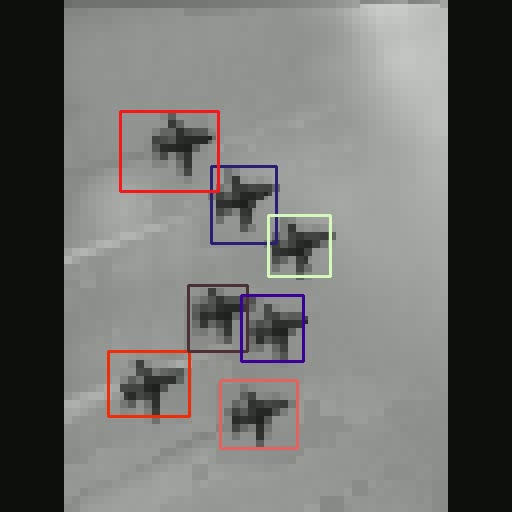}\par\caption*{Mixed NN}
    \includegraphics[width=\linewidth]{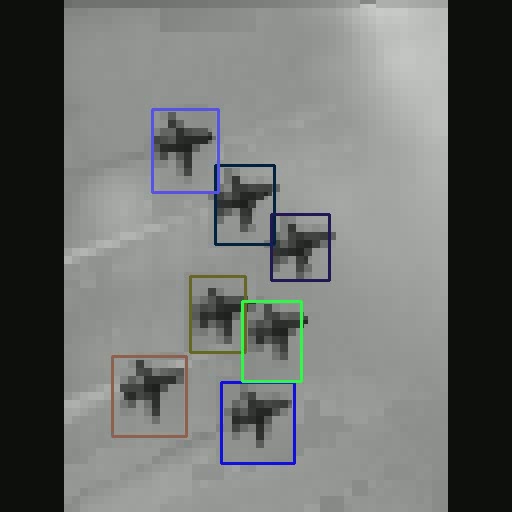}\par\caption*{MixedU NN}
    \includegraphics[width=\linewidth]{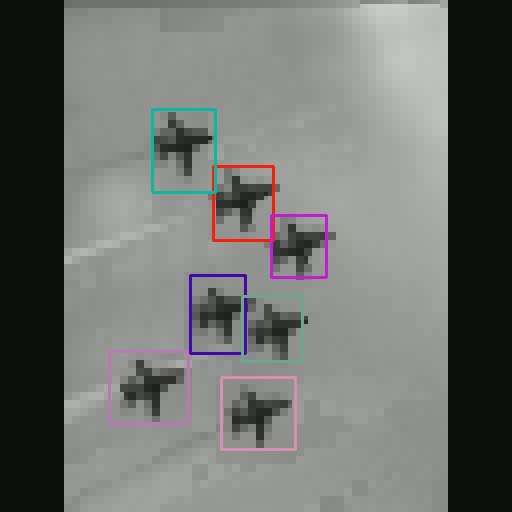}\par\caption*{Mixed+ NN}
    \includegraphics[width=\linewidth]{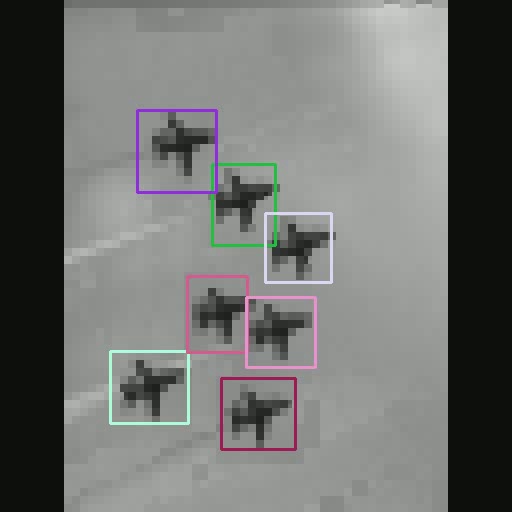}\par\caption*{MixedU+ NN}
\end{multicols}
\begin{multicols}{6}
    \noindent
    \includegraphics[width=\linewidth]{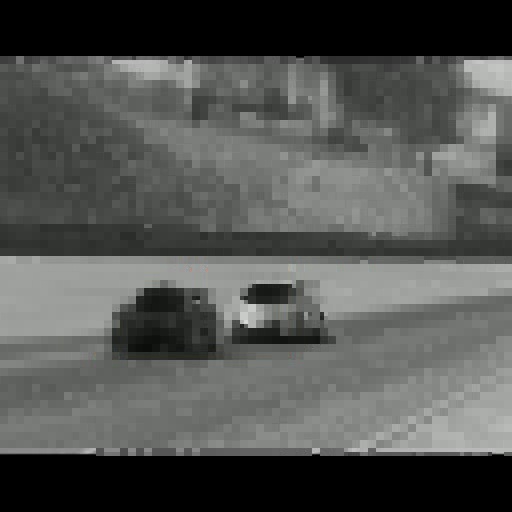}\par\caption*{Pristine NN}
    \includegraphics[width=\linewidth]{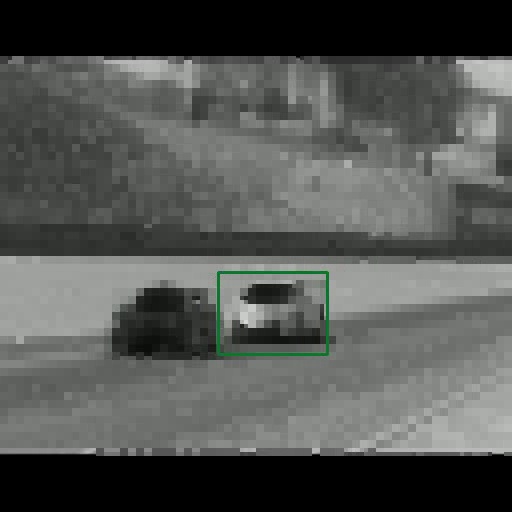}\par\caption*{Uniform NN}
    \includegraphics[width=\linewidth]{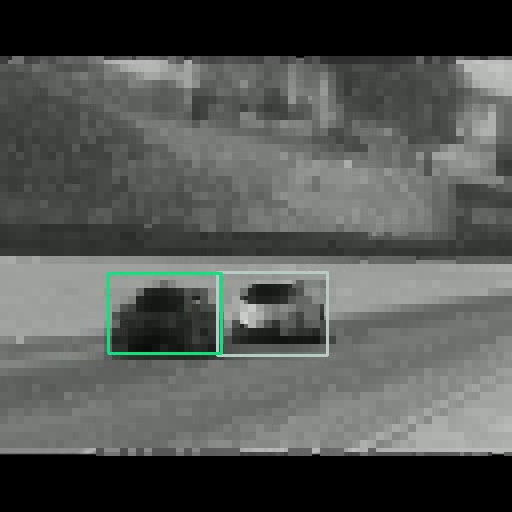}\par\caption*{Mixed NN}
    \includegraphics[width=\linewidth]{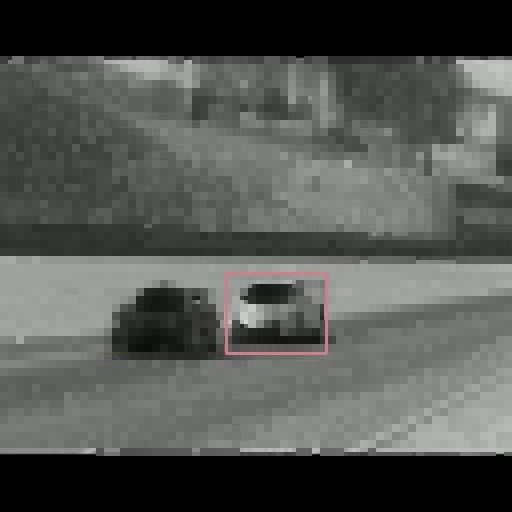}\par\caption*{MixedU NN}
    \includegraphics[width=\linewidth]{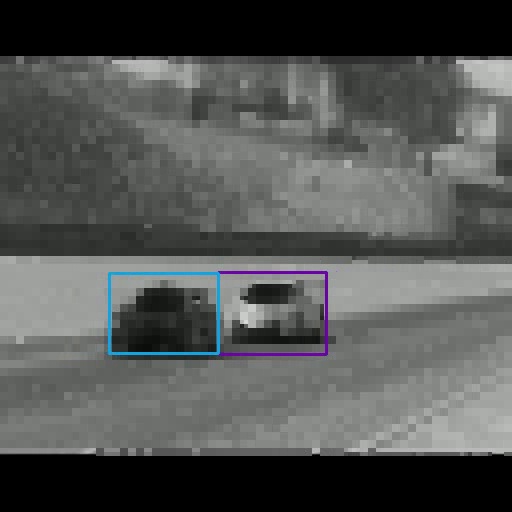}\par\caption*{Mixed+ NN}
    \includegraphics[width=\linewidth]{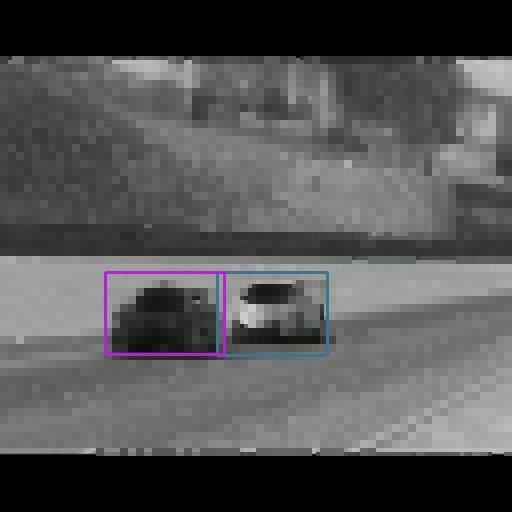}\par\caption*{MixedU+ NN}
\end{multicols}
\begin{multicols}{6}
    \noindent
    \includegraphics[width=\linewidth]{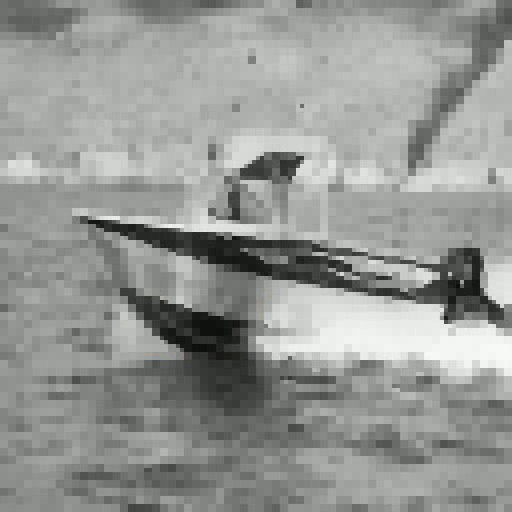}\par\caption*{Pristine NN}
    \includegraphics[width=\linewidth]{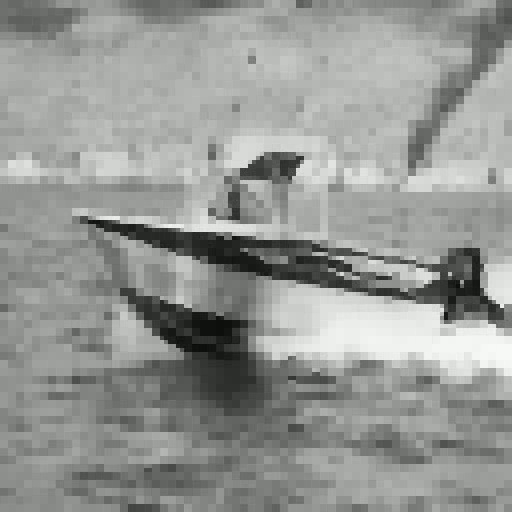}\par\caption*{Uniform NN}
    \includegraphics[width=\linewidth]{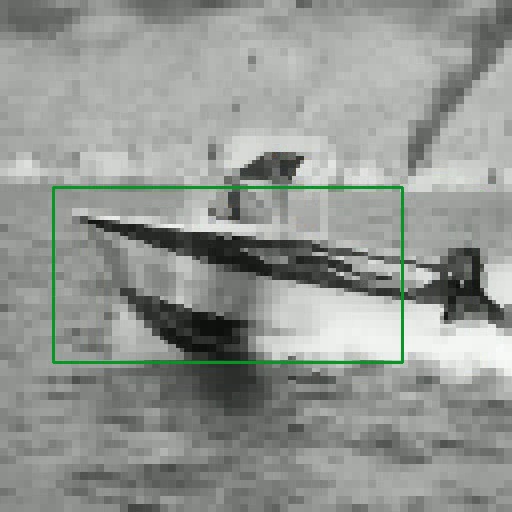}\par\caption*{Mixed NN}
    \includegraphics[width=\linewidth]{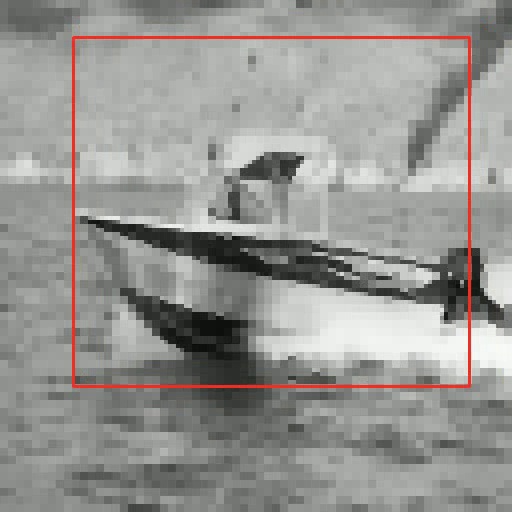}\par\caption*{MixedU NN}
   \includegraphics[width=\linewidth]{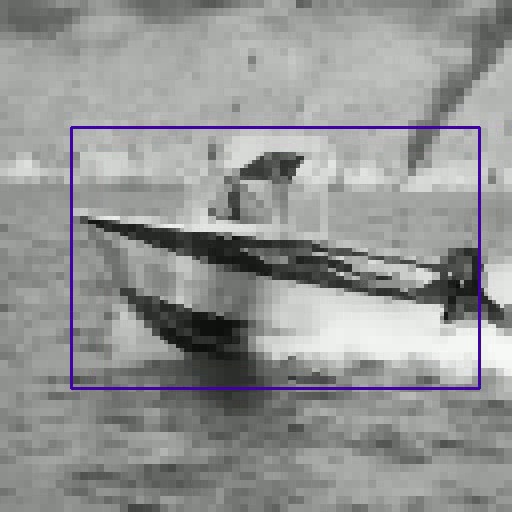}\par\caption*{Mixed+ NN}
    \includegraphics[width=\linewidth]{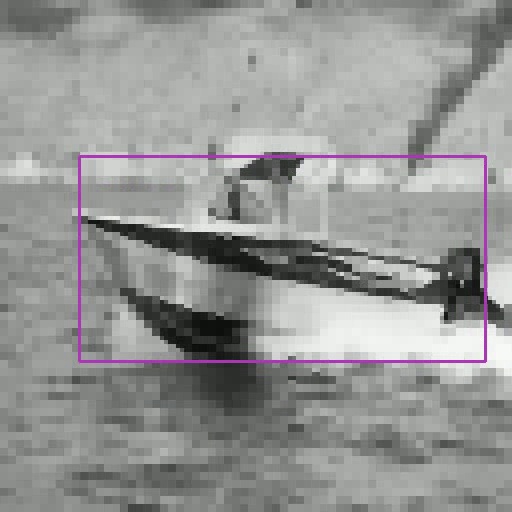}\par\caption*{MixedU+ NN}
\end{multicols}
\caption{Top row: $\hat{f}_{75}$ of the Airplane Sequence, middle row: $\hat{f}_{45}$ of the Car Sequence, bottom row: $\hat{f}_{35}$ of the Watercraft Sequence. Degraded frames generated at $\lambda = 400$. We ask the reader to zoom in on each of the frames to see the degradations.}
\label{fig_comparison}
\end{figure*}

In this mode of operation, $\lambda$ is kept constant. This fluctuates the rate and distortion per frame. MOTA is computed for each sequence for the performance of the system trained with different object detectors. The effect of 2 step-training methodology is demonstrated here. Tracker assisted object detector upscoring is not included in this subsection of experiments. Fig. \ref{fig_comparison} show the detections in the distorted frame of airplane, car and watercraft sequence for each of the six Faster R-CNN models, with distorted frames generated at $\lambda=400$. The Pristine NN detector fails to detect the objects in each of the three cases. On the other hand, the Uniform NN detector detects a few objects. Mixed NN and MixedU NN detectors are able to detect almost all of the objects. However, the bounding boxes given by these detectors either overfit the objects with excess background included or underfit the objects. On the other hand Mixed+ and MixedU+ NN detectors perform a better job in fitting the bounding box to the objects in the scene including minimum background.

\begin{figure*}[t]
\begin{multicols}{3}
    \noindent
    \includegraphics[width=\linewidth,height=125pt]{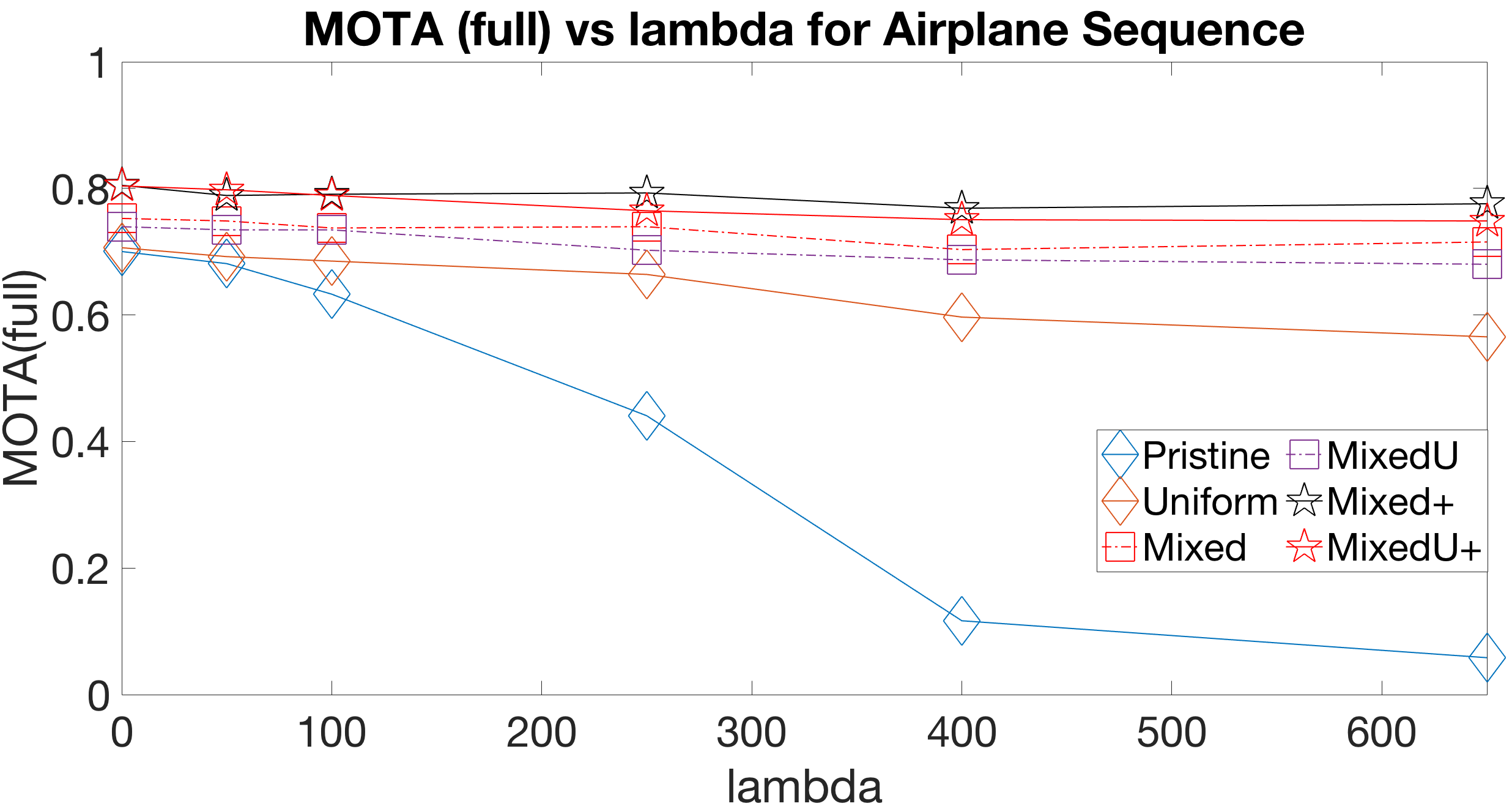}\par\caption*{(a) Airplane $MOTA_{full}$}
    \includegraphics[width=\linewidth, height=125pt]{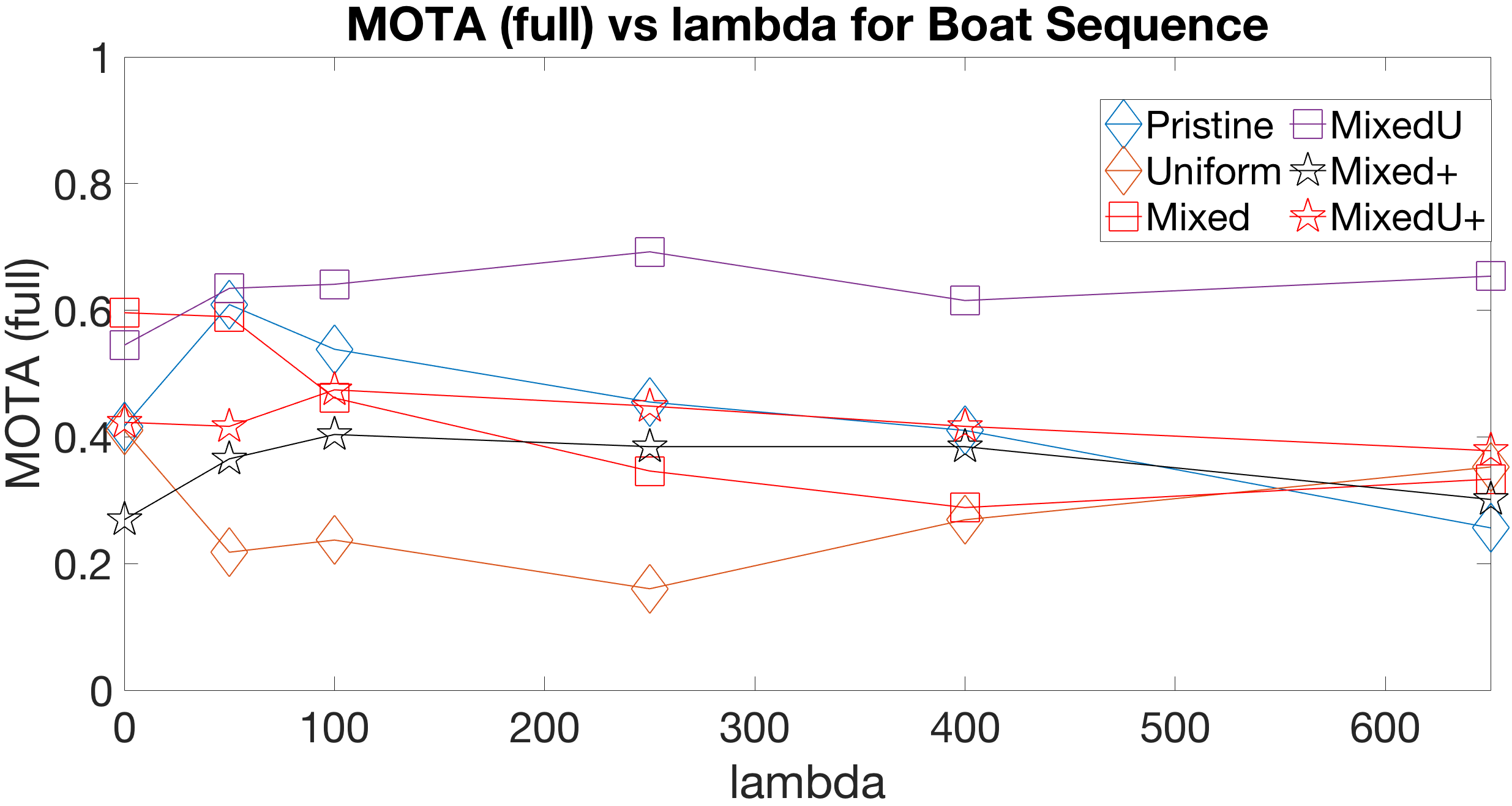}\par\caption*{(b) Watercraft $MOTA_{full}$}
    \includegraphics[width=\linewidth, height=125pt]{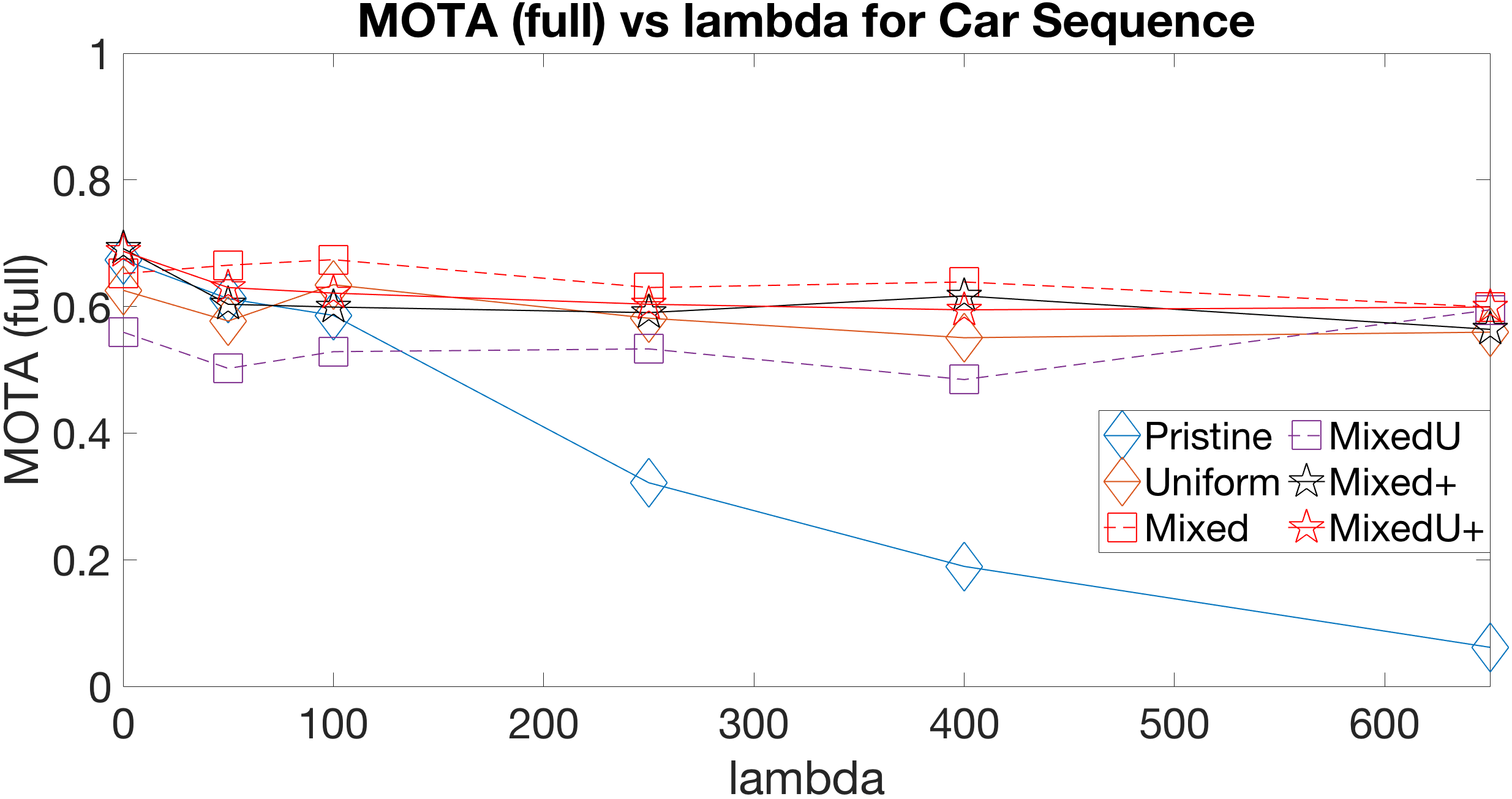}\par\caption*{(c) Car $MOTA_{full}$}
\end{multicols}
\caption{$MOTA_{full}$ vs $\lambda$ for Airplane, Car and Watercraft sequence}
\label{fig_mota_full_lambda_comparison}
\end{figure*}

\begin{figure*}[t]
\begin{multicols}{3}
    \noindent
    \includegraphics[width=\linewidth,height=125pt]{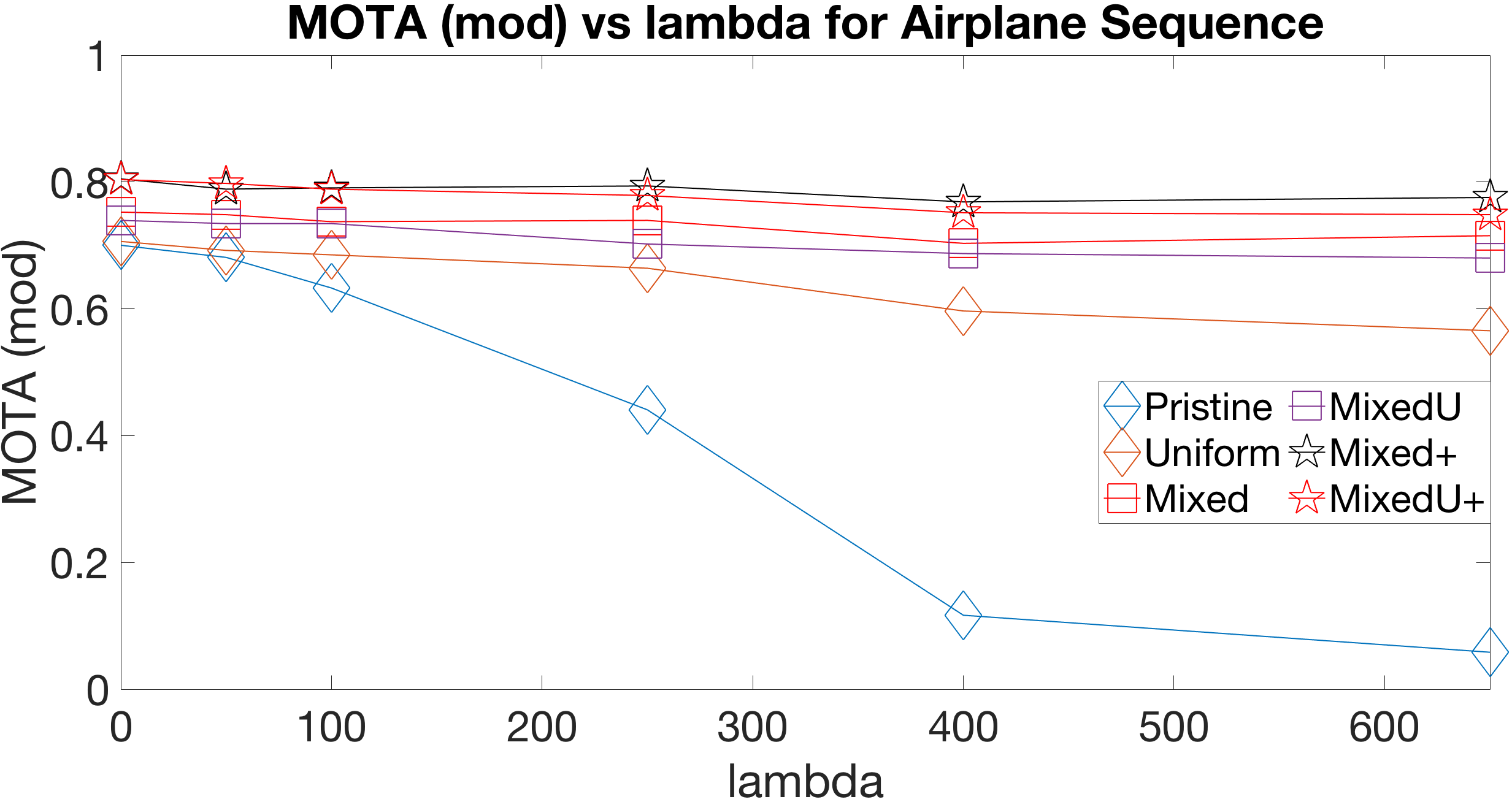}\par\caption*{(a) Airplane $MOTA_{mod}$}
    \includegraphics[width=\linewidth, height=125pt]{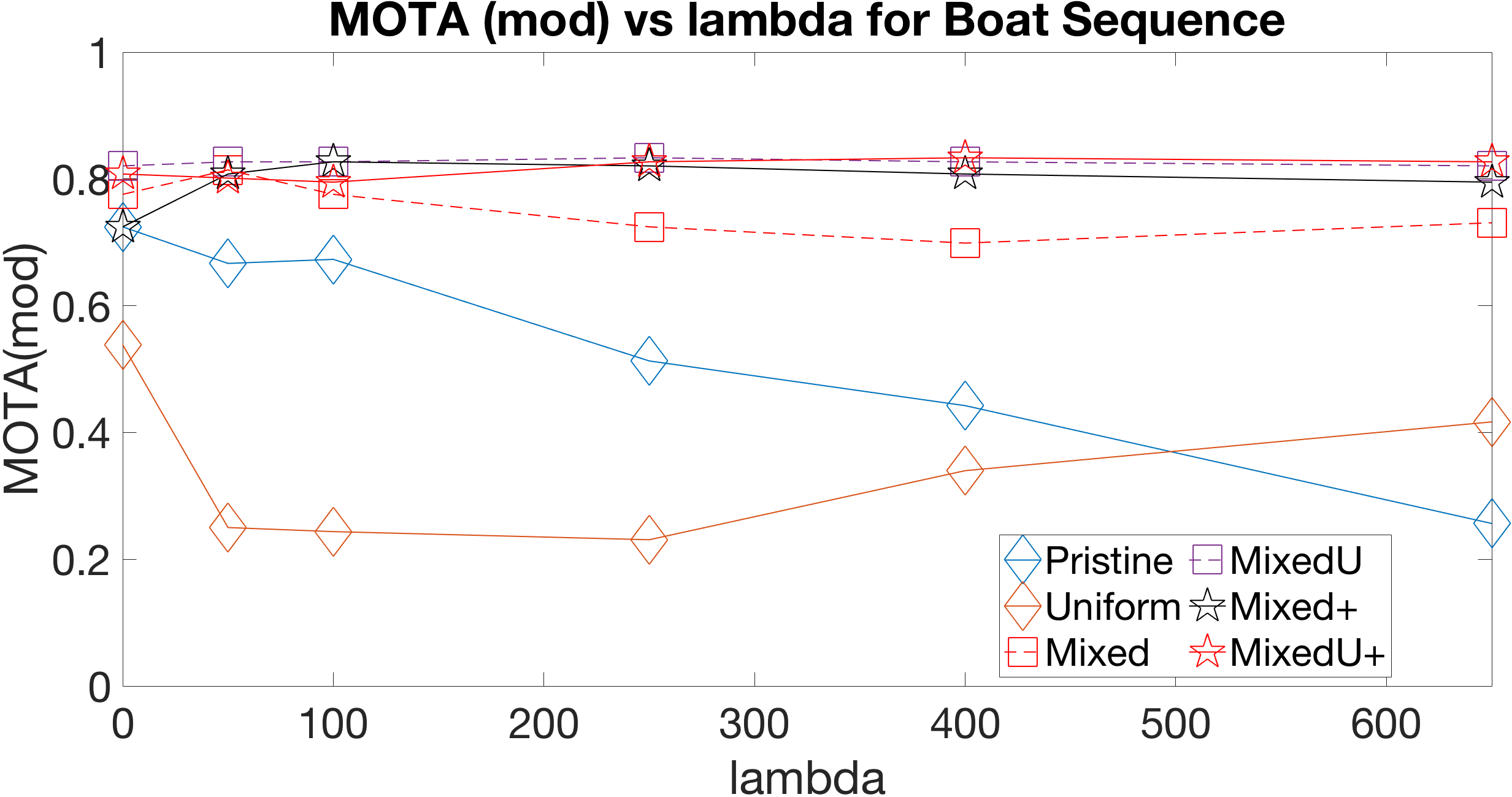}\par\caption*{(b) Watercraft $MOTA_{mod}$}
    \includegraphics[width=\linewidth, height=125pt]{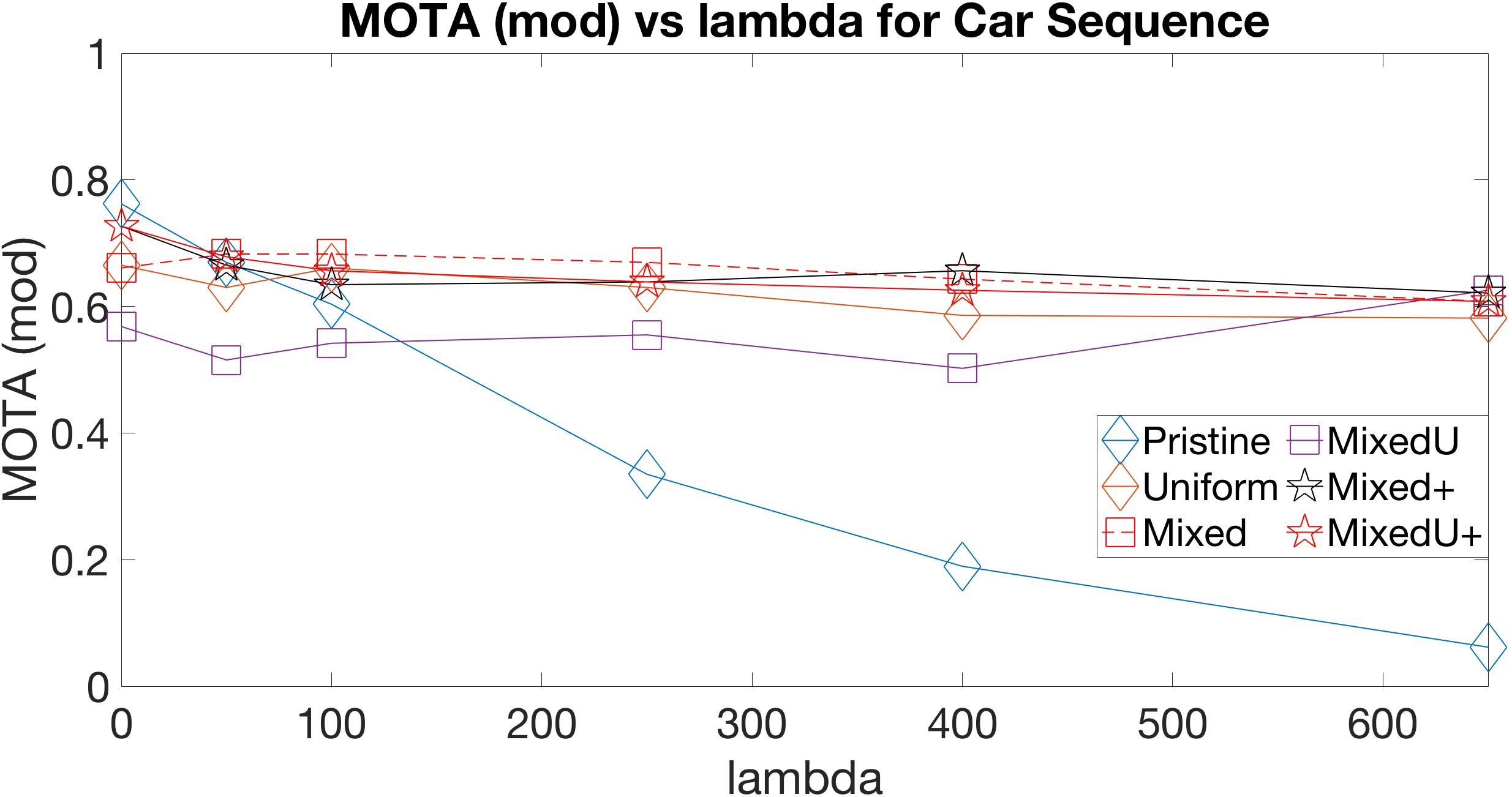}\par\caption*{(c) Car $MOTA_{mod}$}
\end{multicols}
\caption{$MOTA_{mod}$ vs $\lambda$ (Eq. $10$) for Airplane, Car and Watercraft sequence}
\label{fig_mota_mod_lambda_comparison}
\end{figure*}

Fig. \ref{fig_mota_full_lambda_comparison} (a) and \ref{fig_mota_mod_lambda_comparison} (a) shows the $MOTA_{full}$ and $MOTA_{mod}$ curves for the airplane sequence considering Eqns. $9$ and $10$ respectively. There are seven small sized airplanes within the full frame, and some become obscured over time. It is seen that the system trained with the Pristine NN has significant deterioration in performance after $\lambda=250$. The performance of the Uniform NN is significantly better for higher $\lambda$ values than the Pristine NN detector. This indicates that the QT for small objects can be replaced with uniform binning. However, the performance of Mixed and MixedU NN-based detectors is better than the Uniform NN detector---clearly suggesting the benefits of using actual degraded data generated by the system for training the Faster R-CNN. The best performance is obtained by using the Mixed+ and MixedU+ NNs. The Mixed+ NN detector performance is slightly better than the MixedU+ NN detector since the exact degradations correspond to the QT binning. In MixedU+ NN, the NN has been trained on actual system generated distortions as well as uniformly binned data. Thus the 2 step training strategy do help in improving the performance metric.

The performance of the system when tested on medium-sized cars is shown in the $MOTA_{full}$ and $MOTA_{mod}$ curves of Fig. \ref{fig_mota_full_lambda_comparison} (c) and \ref{fig_mota_mod_lambda_comparison} (c) curves. It is seen here as well that the Pristine NN has a performance drop after $\lambda = 250$. The Uniform NN detector performance is better for higher $\lambda$ values compared to the Pristine NN detector. The Mixed NN detector has higher accuracy than the Uniform NN detector. The MixedU NN detector performance is worse than the Mixed NN and Uniform NN detectors. This indicates that training using both system generated and uniform distortions may lead to a sub-optimal performance. However when we do Step-II training, the performance of MixedU+ is greater than MixedU. The Mixed+ NN detector performance is within about 0.05 at worse ($\lambda=100$) to the Mixed NN detector for most of the $\lambda$ values. The Mixed NN detector trained with only the system generated data once, has its performance close of the Mixed+ NN and MixedU+ NN detector. However, overall, the performance of MixedU+ and Mixed+ NN detectors is better than that of the pristine detectors.

The watercraft sequence has a large boat which occupies most of the frame during the entire sequence. The performance of the system for this sequence is shown in Fig. \ref{fig_mota_full_lambda_comparison}(b) and \ref{fig_mota_mod_lambda_comparison}(b). The Pristine NN performance drops significantly beyond $\lambda \geq 250$ as in the previous two cases of small and medium sized objects. The Uniform NN detector performance is lower for most of the $\lambda$ values than the Pristine NN detector. The Mixed NN detector and MixedU NN detector performance is higher than the Pristine NN detector. Surprisingly, the MixedU NN detector performance is higher than the Mixed NN detector's performance. This implies that for large sized objects, the system generated distortion is different from uniform binned distortions, and training the detector with both these types of distortions actually aids the performance. The performance of Mixed+ NN detector is better than Mixed NN detector. However, the performance of MixedU NN detector is higher due to less false positives. Considering no false positives in our metric, the peformance of MixedU+ NN, MixedU NN and Mixed+ detectors are very similar. The MixedU+ detector performance is within about 0.05 at worse ($\lambda=100$) to the MixedU NN detector for most of the $\lambda$ values. In this case, the 2-step training process does improve the system performance especially when we use the $MOTA_{mod}$ metric, thus highlighting benefit of this training process.      

From our experimental studies, we observe that Mixed and MixedU detectors is able to peform better for medium and large sized objects respectively, mostly due to lack of false positives. However, the performance of the Mixed+ and MixedU+ detectors are the best among the different Faster R-CNN models across the board especially when we ignore false positives. It is also observed that when background objects are significantly present (in the boat sequence),  MixedU and Mixed NN detector tends to perform better with false positives considered in $MOTA_{full}$. However, the experimental studies suggest the benefits of 2-step training process for improving the performance metric for most of the cases. The object detector trained only once (MixedU and Mixed) has performance improvements over Pristine NN detector as well but in general the performance gains are lower than that of the 2-step trained MixedU+ and Mixed+ models.  

\subsection{Operation at Constant Bit Rate}
\begin{figure*}[t]
\begin{multicols}{3}
    \noindent
    \includegraphics[width=\linewidth, height=125pt]{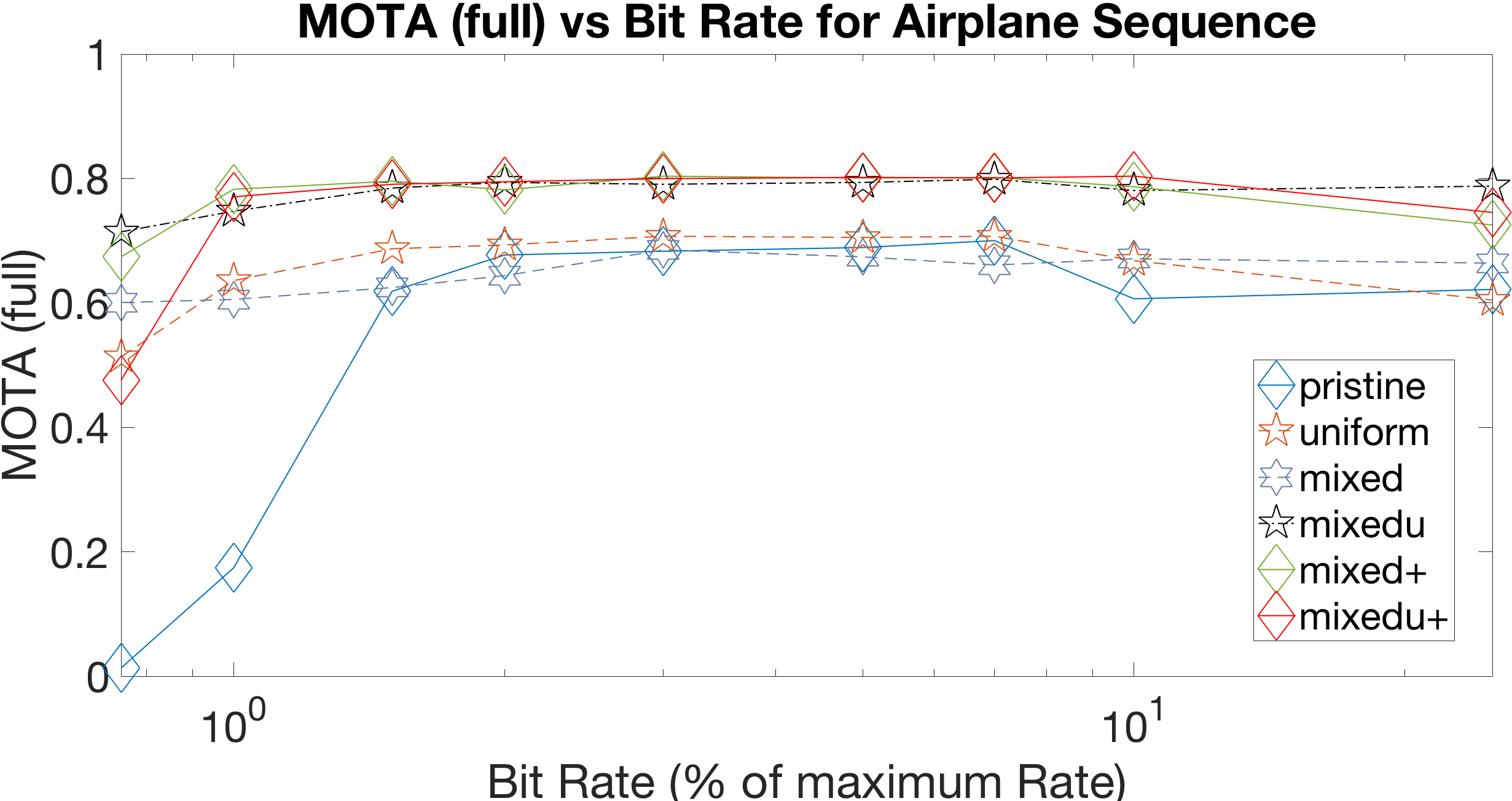}\par\caption*{(a) Airplane $MOTA_{full}$}
    \includegraphics[width=\linewidth, height=125pt]{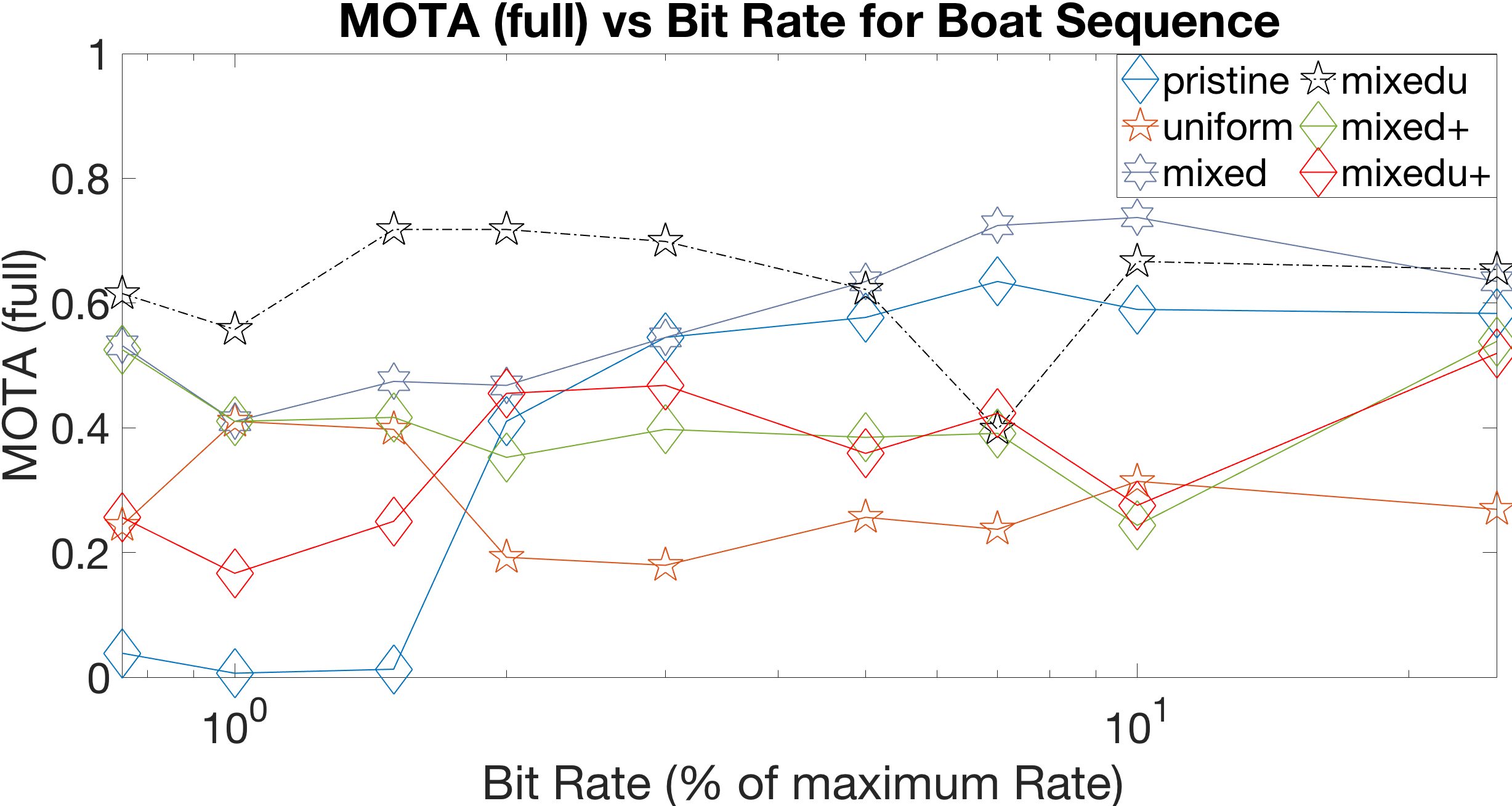}\par\caption*{(c) Watercraft $MOTA_{full}$}
    \includegraphics[width=\linewidth, height=125pt]{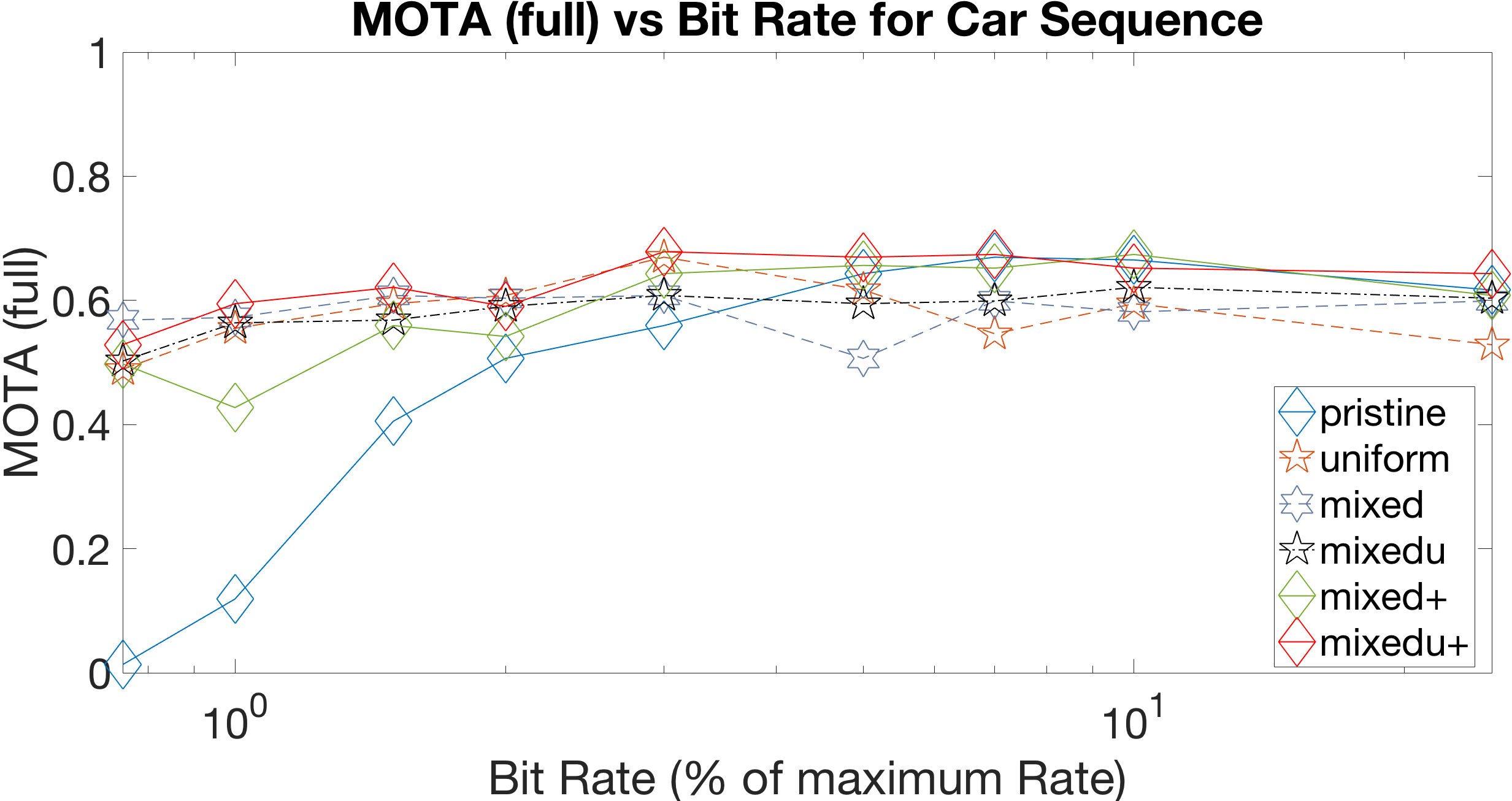}\par\caption*{(c) Car $MOTA_{full}$}
\end{multicols}
\caption{$MOTA_{full}$ vs Rate Curves for Pristine, Uniform, Mixed, MixedU, Mixed+ and MixedU+ detectors.}
\label{fig_ratec_comparison_full_all}
\end{figure*}

\begin{figure*}[t]
\begin{multicols}{3}
    \noindent
    \includegraphics[width=\linewidth, height=125pt]{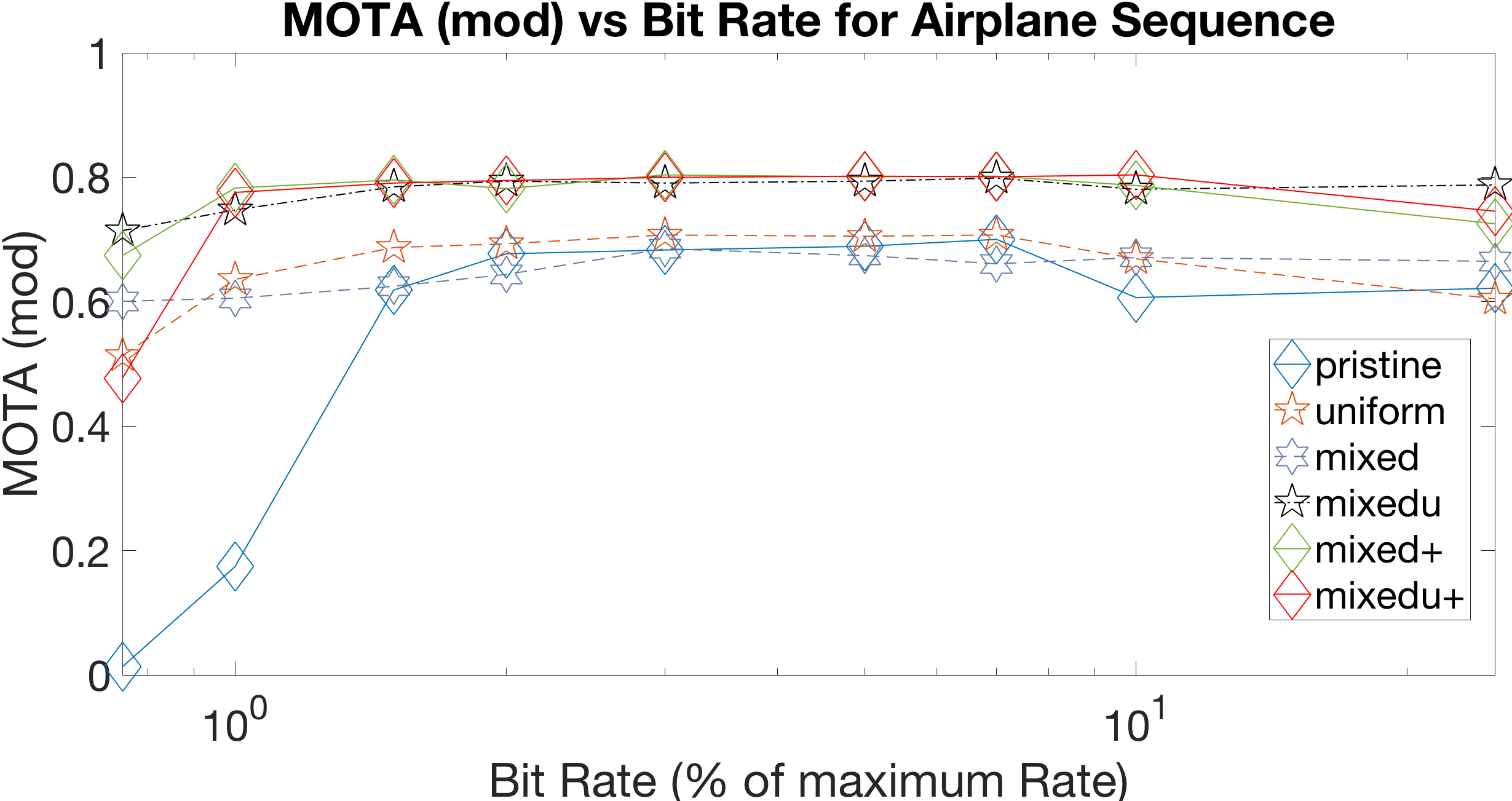}\par\caption*{(a) Airplane $MOTA_{mod}$}
    \includegraphics[width=\linewidth, height=125pt]{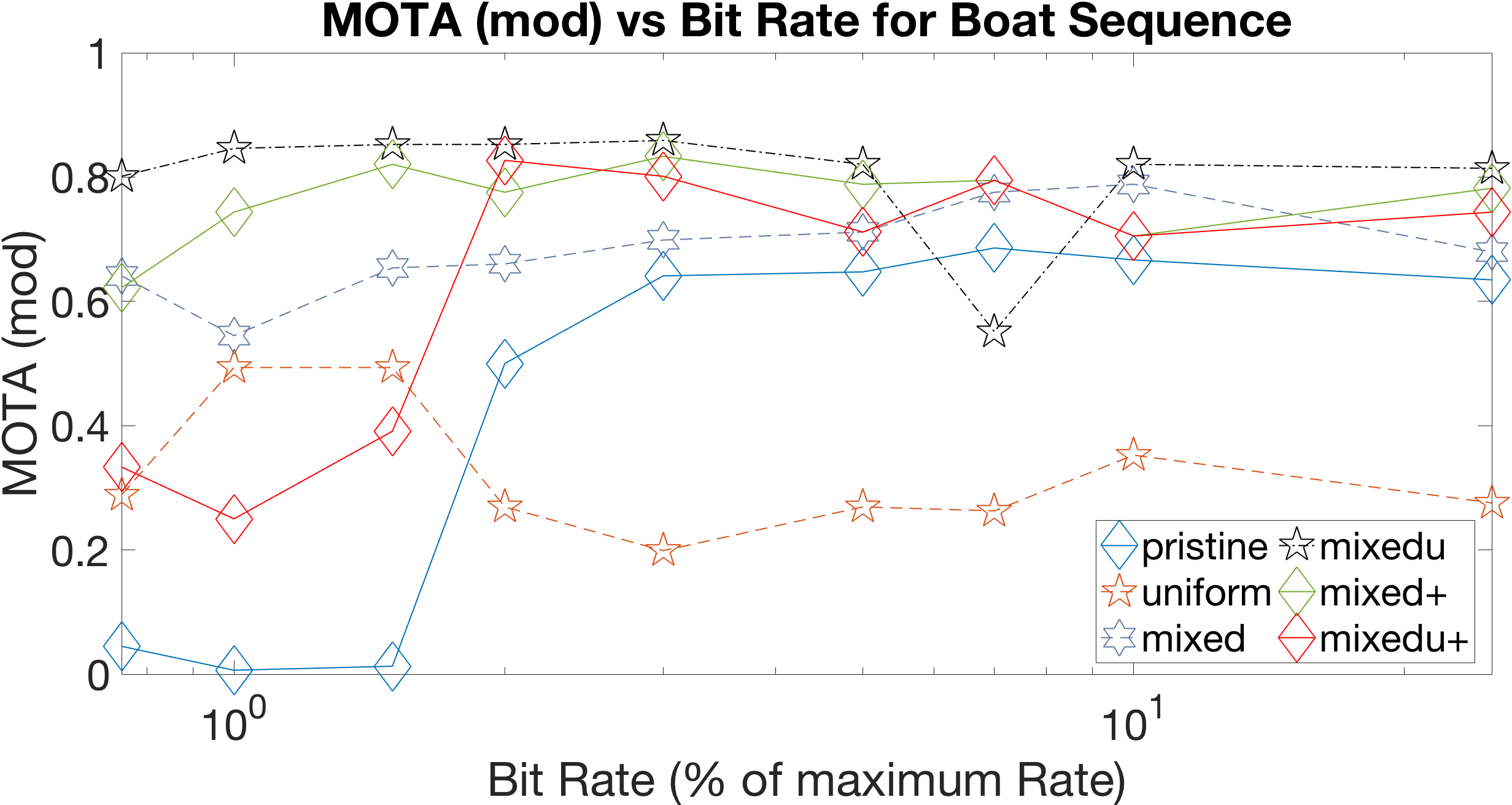}\par\caption*{(c) Watercraft $MOTA_{mod}$}
    \includegraphics[width=\linewidth, height=125pt]{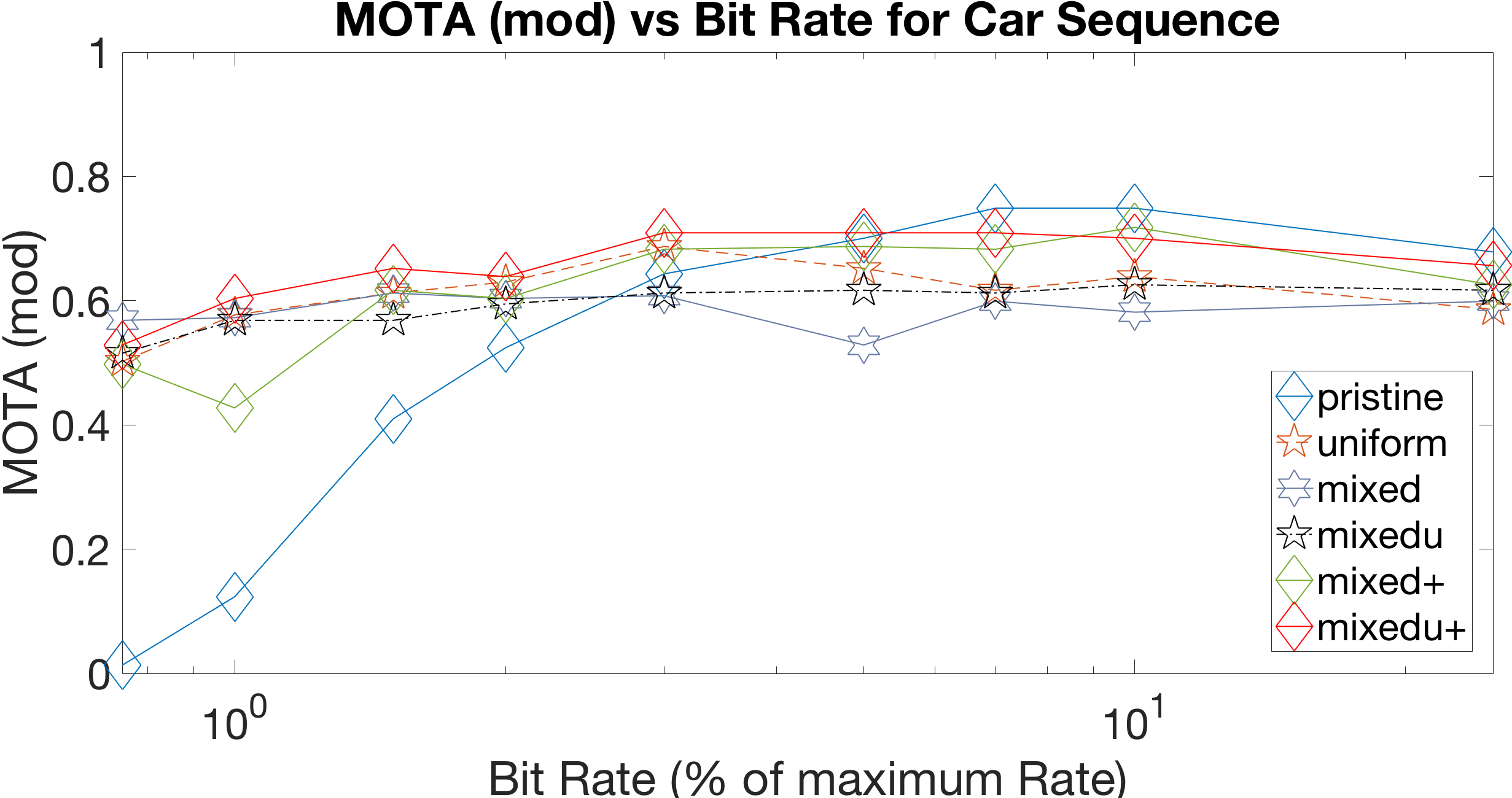}\par\caption*{(c) Car $MOTA_{mod}$}
\end{multicols}
\caption{$MOTA_{mod}$ vs Rate Curves for Pristine, Uniform, Mixed, MixedU, Mixed+ and MixedU+ detectors.}
\label{fig_ratec_comparison_mod_all}
\end{figure*}

In the mode of operation for the system, we force the bit rate to be constant as a fraction of the maximum bit rate (within a tolerance of $1 \% $ of the fractional bit rate). This makes $\lambda$ and the distortion fluctuate in each frame and in each sequence. The detector has been trained with the 2-step strategy. Tracker  assisted  object  detector upscoring  is  not  included  in  this  subsection  of  experiments.  $MOTA_{full}$ and $MOTA_{mod}$ is computed for each of these rates. Fig. \ref{fig_ratec_comparison_full_all} and \ref{fig_ratec_comparison_mod_all} shows the plot of $MOTA_{full}$ and $MOTA_{mod}$ vs bit-rate as per Eqns. $9$ and $10$ respectively. We have computed the performance using the Pristine, Uniform, Mixed, MixedU, Mixed+ and MixedU+ NN detectors to show their overall performance with each detector. Both $MOTA_{full}$ and $MOTA_{mod}$ increases initially with the increase in the bit rate for the Airplane, Watercraft and Car sequences and then remains approximately constant. The false positives are very few as the $MOTA_{full}$ and $MOTA_{mod}$ values are close to each other. The performance of the Mixed+, MixedU+ and MixedU NN detectors are close to each other, with MixedU NN detector performance having less false positives. However, across the board, the Mixed+ detector have consistent good performance than MixedU+ detector.

It is also pointed out that for the watercraft sequence especially at lower bit rates ($<2\%$), in some frames we have values of $\lambda$ well over the maximum $\lambda = 650$, the maximum $\lambda$ we had trained the detectors. Yet, the system trained at medium distortions can even perform quite well at these higher distortions. This shows the robustness of the 2-step training process at distortion levels worse than the trained distortion levels. The early convergence of the curves to high $MOTA_{full}$ and $MOTA_{mod}$ accuracy at low bit rates show the effectiveness of the 2-step training procedure over using a Pristine NN detector. The system performance has been shown for 0.75 \% to 25 \% of the maximum bit rate of 62.9 Mbits/s which is the desired range of operation.

\subsection{Operation with Tracker assisted Detection Framework}
\begin{figure*}[t]
\begin{multicols}{3}
    \noindent
    \includegraphics[width=\linewidth,height=125pt]{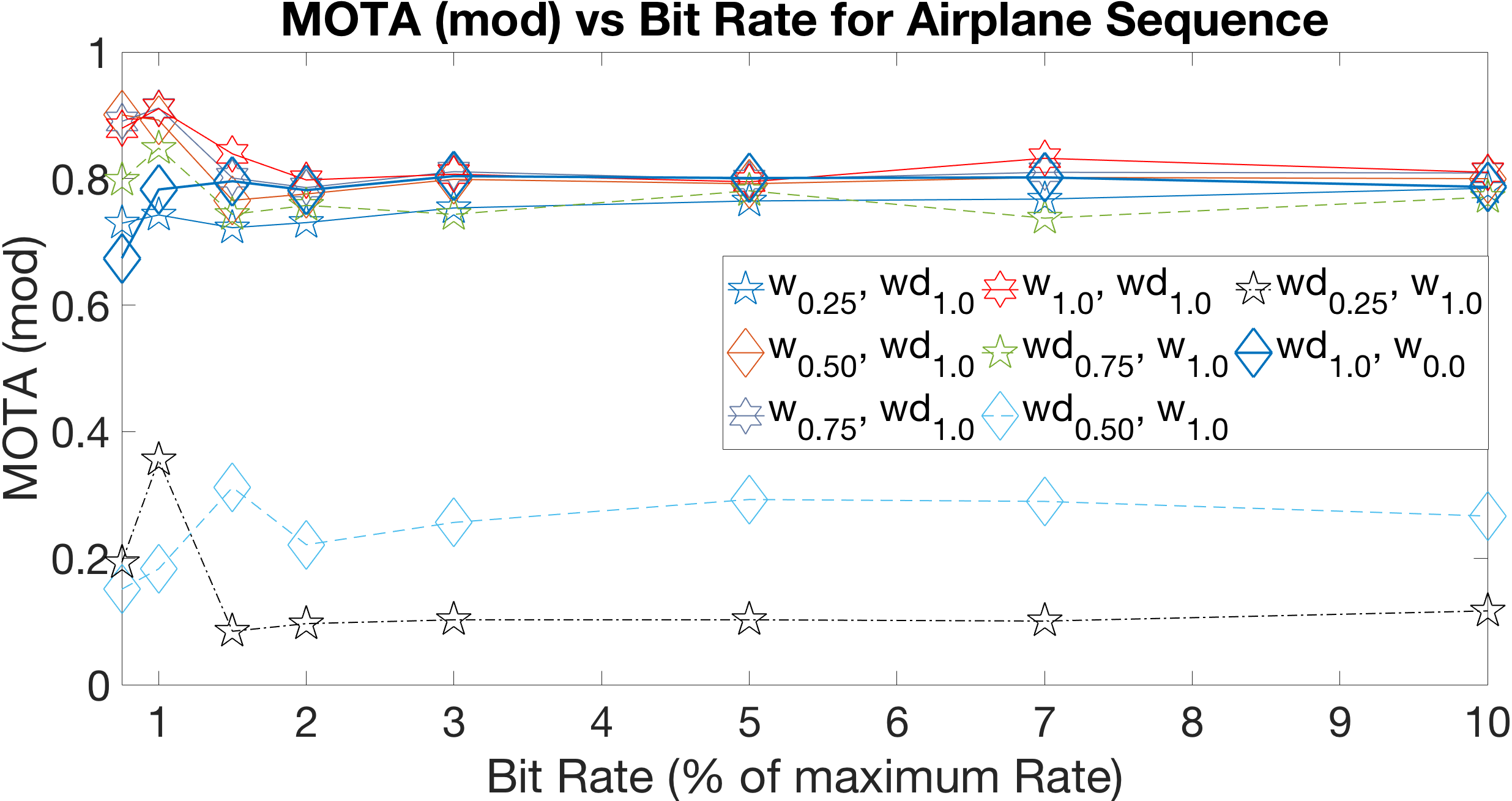}\par\caption*{(a) Airplane $MOTA_{mod}$}
    \includegraphics[width=\linewidth,height=125pt]{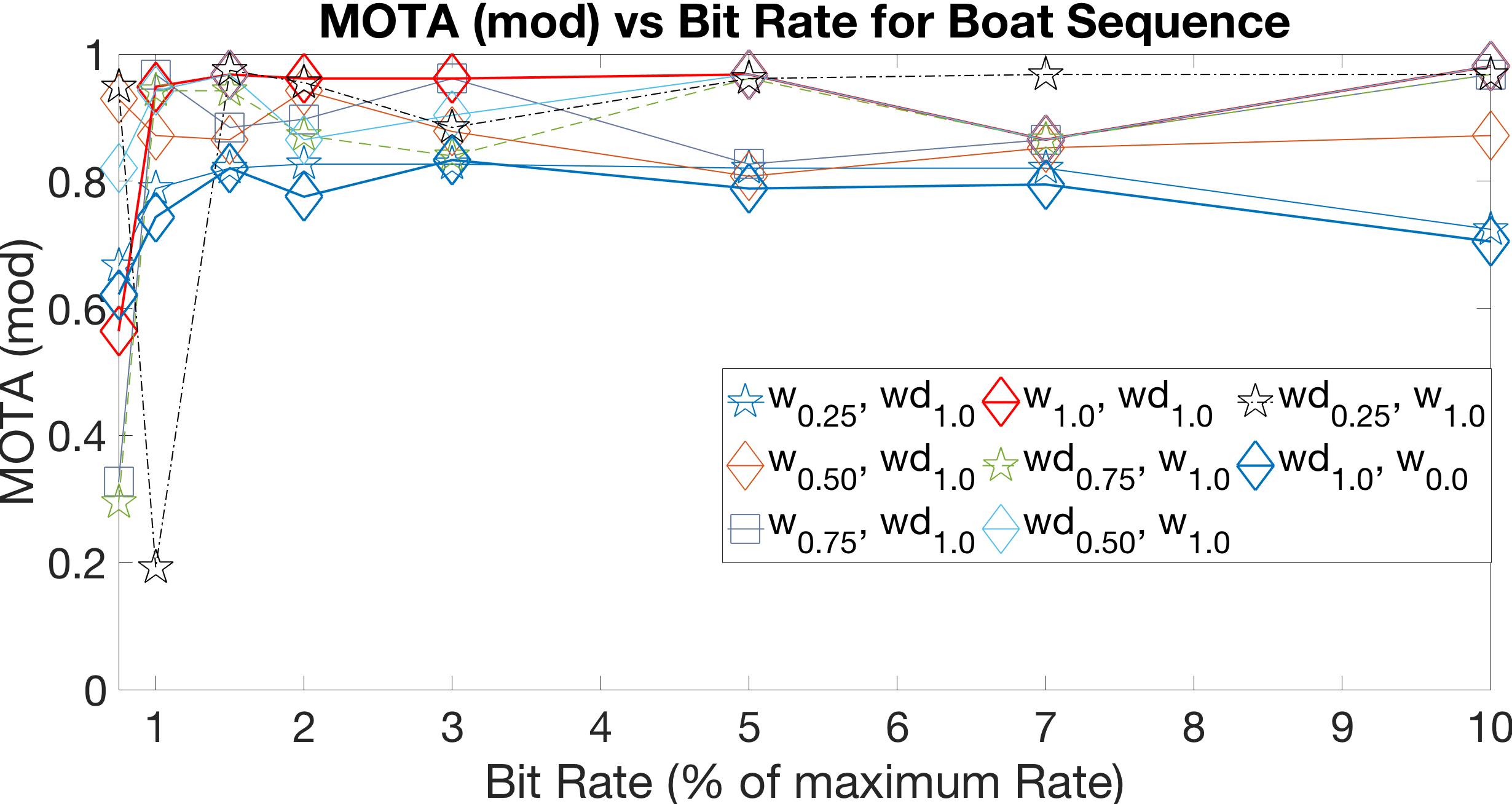}\par\caption*{(b) Watercraft $MOTA_{mod}$}
    \includegraphics[width=\linewidth,height=125pt]{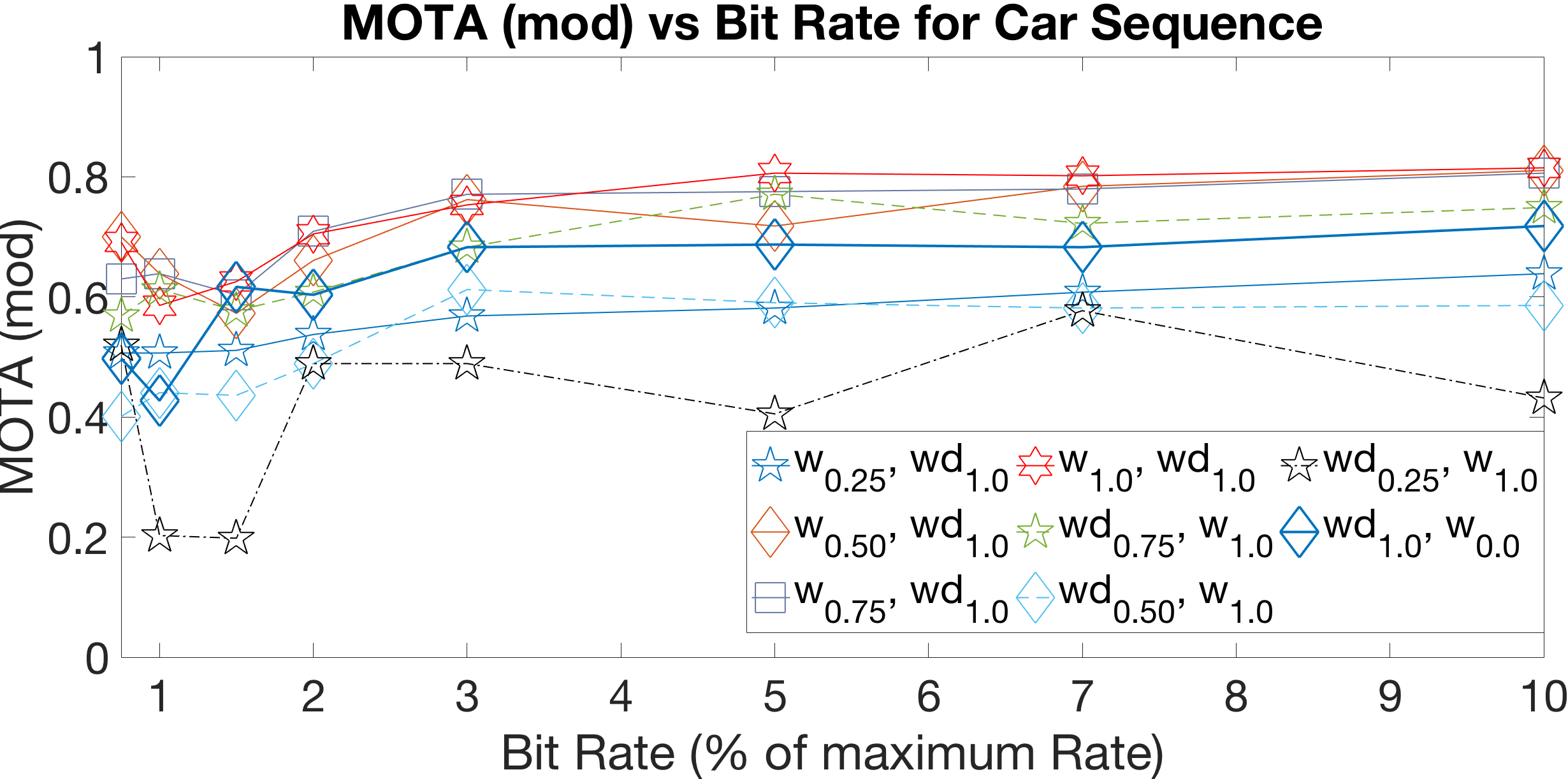}\par\caption*{(c) Car $MOTA_{mod}$}
\end{multicols}
\caption{System performance with varying $w_d$ and $w_t$ with Mixed+ detector for the airplane, watercraft and car sequences with Tracker assisted Detector.}
\label{Thomas_upscoring}
\end{figure*}

We compare the performance of the system with tracker assisted object detection alongwith the 2-step training strategy for the object detector as mentioned in the previous section. We do a parametric evaluation of the system performance with varying tunable detection weight $w_d$ and tracking weight $w_t$ as shown in Fig. \ref{Thomas_upscoring}. Mixed+ object detector has been used in the experiments as it provides one of the best performance for the system as shown in the previous subsection. We observe that for fixed $w_t$ (referred in Fig. \ref{Thomas_upscoring} as $w$), the performance detoriates with reduction in $w_d$ (referred in Fig. \ref{Thomas_upscoring} as $wd$) in most of the cases. On the other hand for a fixed $w_d$, the performance of the system is better when $w_t$ is increased. Based on our experimental results, we find the best performance $MOTA_{mod}$ in most of the cases is when $w_d = 1$ and $w_t = 1$. It is clearly evident from our experiments that there is a significant increase in the system peformance when we have a object detector assisted with the tracker compared to the system with no assistance from the object tracker $w_d = 1$ and $w_t = 0$, especially when there is significant background, as in the boat and car sequences.

\subsection{Comparison with other methods}
We compare the performance of our method with three other techniques. One of the alternative compression techniques is simple binning of images (without using our system) to $2\times2$, $4\times4$, $8\times8$ and $16\times16$ blocks with each block having the intensity value equal to the average of individual pixels within the block. In the case of uniformly binned frames, the pristine detector is used to evaluate the MOTA metric. Alternatively, the video is separately compressed using sophisticated H.264 (AVC) and H.265 (HEVC) techniques, which are most commonly used video compression standards in the video and telecom industry. We utilize FFmpeg library libx265 with its HEVC video encoder wrapper (x265). Similarly for H.264 compression FFmpeg library libx264 is used. We use $2-$pass encoding scheme for both H.264 and H.265 as the rate control mode to limit the bit-rate. For fair comparison, we compute the performance metric at the same bit-rates of $0.39 \%$, $1.5 \%$, $6.25 \%$ and $25 \%$ of the maximum bit-rate which is identical to $1/256$, $1/64$, $1/16$ and $1/4$ of the maximum bit-rates respectively. The performance $MOTA_{mod}$ of the videos compressed with naiive binning, AVC and HEVC standarads has been evaluated with pristine object detectors. These compression standards compress videos with high PSNR and high quality. This makes it more reasonable to use pristine object detector for fair comparison. In our proposed system, we use the Mixed+ and Mixedu+ detectors assisted with the tracker.

\begin{figure*}[t]
\begin{multicols}{3}
    \noindent
    \includegraphics[width=\linewidth, height=125pt]{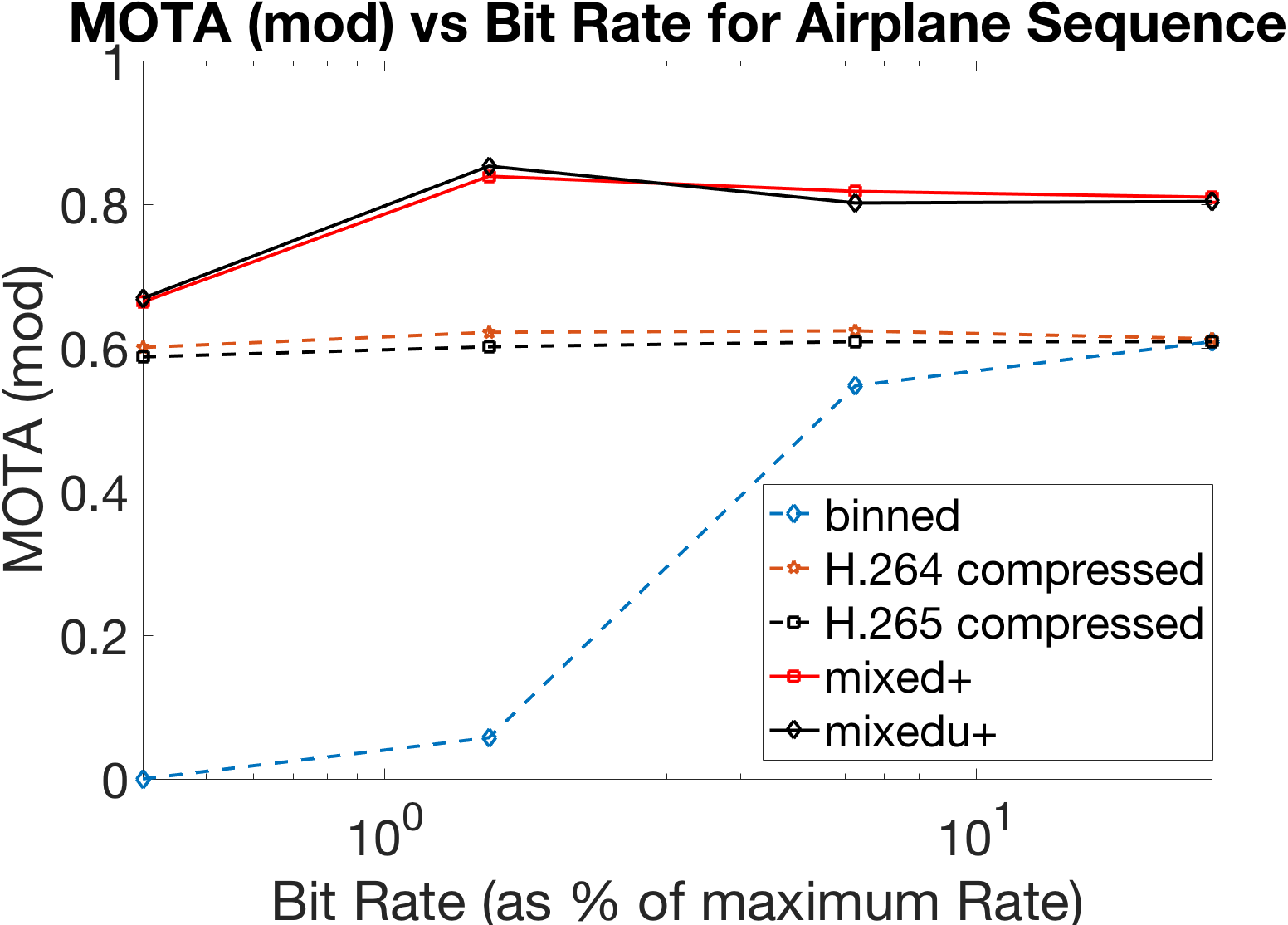}\par\caption*{(a) Airplane $MOTA_{mod}$}
    \includegraphics[width=\linewidth, height=125pt]{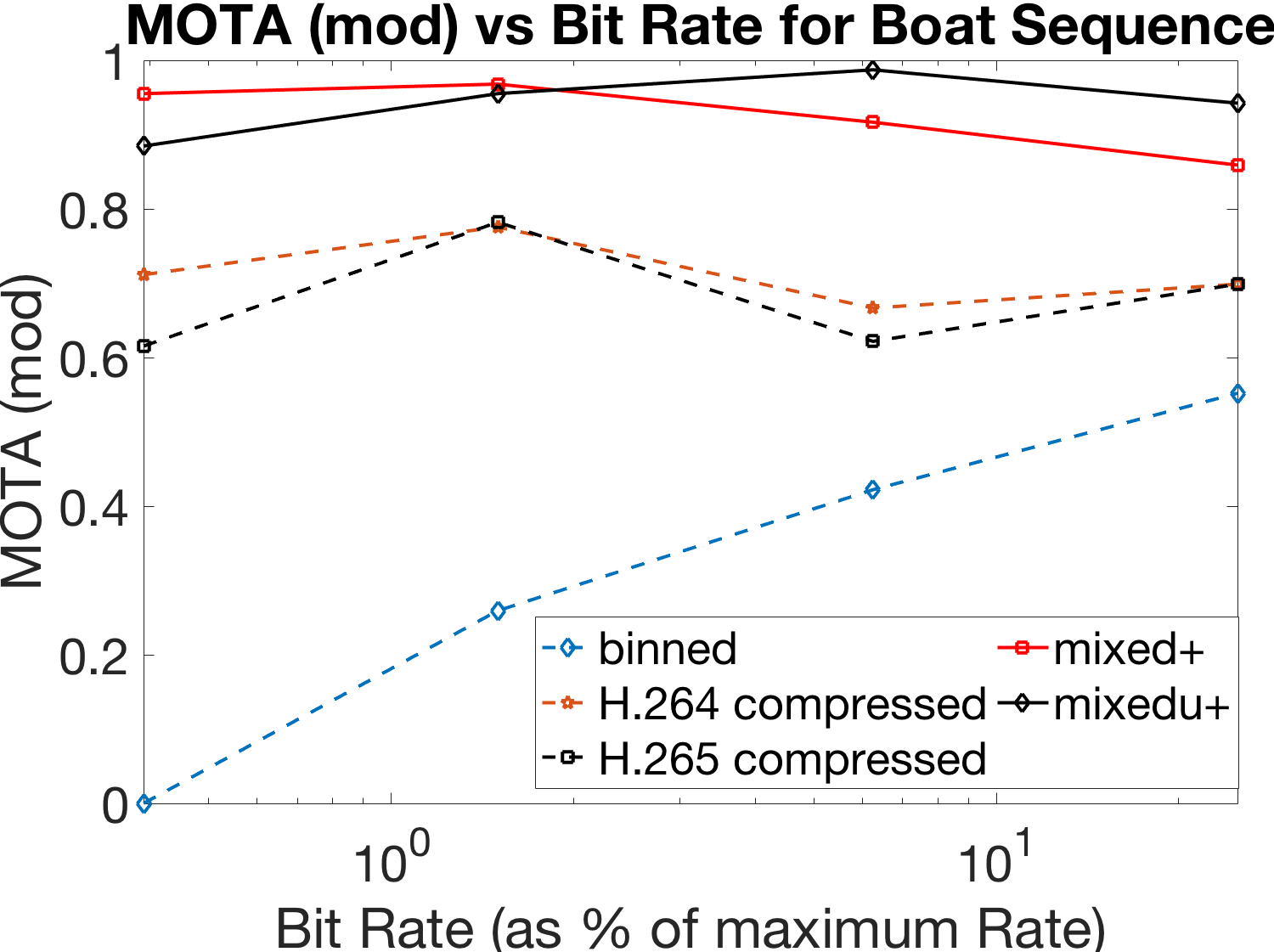}\par\caption*{(b) Watercraft $MOTA_{mod}$}
    \includegraphics[width=\linewidth, height=125pt]{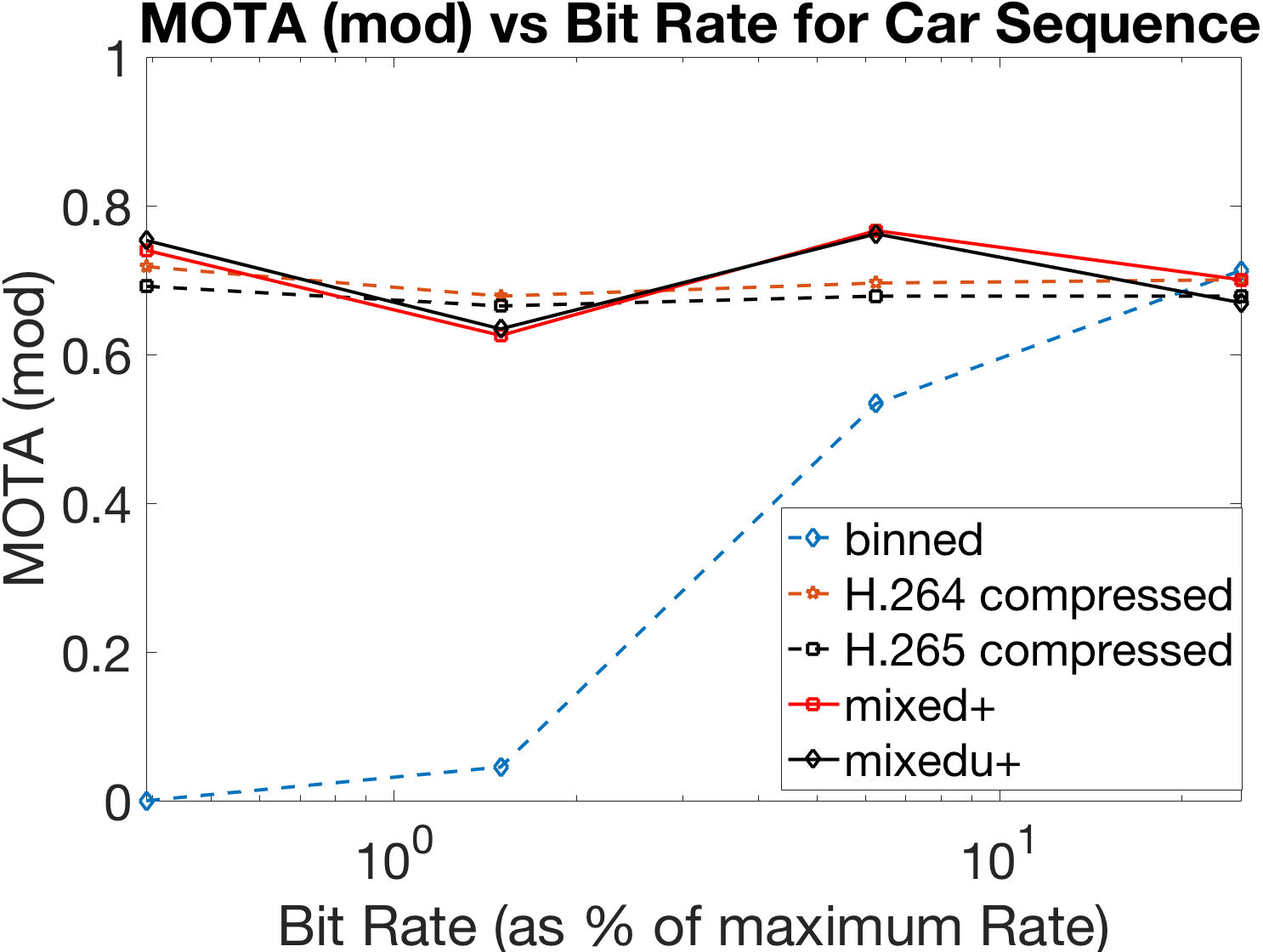}\par\caption*{(c) Car $MOTA_{mod}$}
\end{multicols}
\caption{Comparison of $MOTA_{mod}$ vs Rate Curves (Eq. $10$) for Binned, Mixed+, MixedU+, H.264 and H.265 videos}
\label{comparison_binned_H2645}
\end{figure*}

We see that in Fig. \ref{comparison_binned_H2645}, the performance of naiive simple binning detoriates at rates less than $6 \%$ of maximum rate. On the other hand, performance of our system, H.264 and H.265 compressed videos do not detoriate at lower bit rates. In fact, the MOTA performance of our system is better than H.264 and H.265 encoded videos for most of the cases. It must be kept in mind that sophisticated video coding techniques such as H.264 or H.265 techniques are computationally heavy and is not suitable to be applied directly in a resource constrained chip (as in our current architecture). Thus, with the current computationally constrained chip, our proposed system has good tracking accuracy compared to current state-of-the-art compression standards such as H.264 and H.265.

\section{Discussion} 
In this work, we propose an intelligent algorithm for adaptive sampling of high bit rate data captured by an imager (chip), optimized together with a reconstruction algorithm for object detection and tracking on a remote host. The model has been developed assuming a chip with low computational power and a remote host with high computational power. In this framework, the communication channel between the chip and host has limited bandwidth and thus limited data transfer capabilities. The chip performs the Viterbi optimization for generating QT and skip/acquire modes, while the host perform the tasks of object detection and tracking along with predicting the RoIs in the next time instant for the chip. The performance curves of $MOTA_{full}$ and $MOTA_{mod}$ indicate that the performance of the system deteriorates for the Pristine NN Model beyond $\lambda = 250$. This is consistent among all the categories of objects which have different sizes. It is also evident that the performance of the Faster R-CNN is dependent on the level of QT binning of the ROIs. The edges of the objects get distorted significantly based on the level of QT binning. Additionally, the texture of the object is affected by the QT binning which in turn affects the detector performance. It is consistent with the observation in \cite{Texture} that the ImageNet-trained CNNs are biased towards texture decisions than based on shape.

In our investigation, we find that at high distortions, the background influences the amount of false positives. In the case of a flat background like the airplane sequence, the false positives are fewer. However, this increases in the boat and car sequences which has significant content in the background. The dataset contains small, medium and large sized objects in each class. For high $\lambda$, the distortion is very high and small objects are binned very similarly to the background. This affects the false detections with sufficient background content as the CNN identifies portions of the background as objects. The Faster R-CNN was trained to have a good accuracy over detecting objects of different classes and sizes, which results in more false positives at higher $\lambda$ values that reduce the $MOTA_{full}$ scores. Both $MOTA_{full}$ and $MOTA_{mod}$ scores increases with an increase in bit rate and then saturates. As the rate reduces, the distortion increases. However, both the detectors trained in the 2-step process have their performance at low rates better than Pristine NN detector. Interestingly, the detector trained only once with a mixture of uniformly binned images and system generated images have comparable performance especially over varying bit rates.

We also observe that by adding a tracker assisted object detection on the 2-step training strategy further improves the MOTA. A detailed study on the relative weightage of the detection confidence and tracker confidence proposal bounding boxes have been carried out to find the optimal weights of 1:1 which improves the MOTA scores across the board. The performance of the system is comparable to sophisticated AVC and HEVC techniques which require high computational power on the device. Additionally, our performance metrics is higher than naiive binning techniques especially significantly at lower bit rates.

One limitation of the work is that, there is no sensor-host hardware system developed so far to test out this framework (to the best of authors' knowledge). The gigapixel FPAs currently available in the market lack the ability to allocate bits non-uniformly in different regions of image frame. 

\section{Conclusion}

In summary and conclusion, this paper proposes a novel system using a host-chip architecture for video acquisition optimized for object detection and tracking especially designed for computationally constrained chip (edge devices). Although the system is based on QT compression driven by the ROIs, this architecture is generalizable to other forms of region/block based compression as well. Since QT sub-blocks are inherent in AVC, HEVC and VVC standards, we focus on the optimization of the host-chip architecture based on QT decomposition. Future work will involve exploring other region based compression techniques. A Viterbi-based optimization was used to generate the acquisition modes in the FPA along with the optimal QT structure that minimizes the area-normalized, weighted rate-distortion equation. The optimization algorithm takes into account the priority regions in the scene based on objects of interest. An object detector, Faster R-CNN, is used to detect the ROIs based on the class of object. A novel 2-step training methodology of the Faster R-CNN is applied. In Step-I, we train using ground truth boxes as the output of the detection step to generate training data. In Step-II, we generate new image data using the Step-I detector in our system instead of the ground truth bounding boxes as before. This is done in order to have more realistic data (e.g. imperfect bounding boxes from the last frame affecting the distortion in the present frame). The ROIs from the detector in the current frame are used by the Kalman filter-based tracker to predict the ROIs in the next frame. Another novel tracker assisted upscoring of the object detector has been implemented which aids in further improving the MOTA performance metric. The performance of the system is measured by the $MOTA_{full}$ and $MOTA_{mod}$ scores. The results of our method show significant improvements in the tracking performance and the strength of this host-chip architecture in different operating conditions. Compared to state-of-art highly sophisticated compression techniques employed in image/video coding standards, our system performs better for most of the experimental cases.

\section*{Acknowledgment}

This authors are grateful to Defense Advanced Research Projects Agency (DARPA) for their funding in this project. The work is supported in part by a DARPA Grant No. HR0011-17-2-0044.

\end{document}